\PassOptionsToPackage{usenames,dvipsnames,table}{xcolor} 
\PassOptionsToPackage{outputdir=out}{minted}

\documentclass[sigplan]{acmart}

\pdfoutput=1  %

\acmSubmissionID{14}
\renewcommand\footnotetextcopyrightpermission[1]{}
\usepackage[USenglish]{babel}
\usepackage{pdfrender}
\usepackage{microtype}
\usepackage{endnotes}
\usepackage{extdash}
\usepackage{xspace}
\usepackage{caption}
\usepackage{multirow}
\usepackage{tabularx}
\usepackage{xstring}
\usepackage{color, soul}
\usepackage{enumitem}
\usepackage{multicol}
\usepackage{setspace}
\usepackage{float}
\usepackage{listings}
\usepackage{mdframed}

\setstcolor{red}
\makeatletter                   
\def\mdseries@tt{m}             
\makeatother                    

\usepackage[plain]{fancyref}
\usepackage[draft=true]{minted} 
\usepackage{color}
\usepackage{textgreek}
\usepackage{breakurl}           

\usepackage{amssymb}
\usepackage{lipsum}
\usepackage[final]{pdfpages}
\usepackage{setspace}
\usepackage{epsf}
\usepackage{epsfig}
\usepackage{graphicx}
\usepackage{algorithmic}
\usepackage{amsmath}
\usepackage{letltxmacro}

\usepackage{enumitem}
\usepackage{tikz}
\usepackage{tightenum}
\usepackage{xspace}

\usepackage{url}
\usepackage{multirow}

  \renewenvironment{thebibliography}[1]{%
    \begin{oldthebibliography}{#1}%
      \setlength{\parskip}{0ex}%
      \setlength{\itemsep}{0ex}%
  }%
  {%
    \end{oldthebibliography}%
  }

\newwrite\arxivdeps
\immediate\openout\arxivdeps=\jobname-arxivdeps.log

\newcommand\verifymarkedforarxivfile[1]{%
\ifdefined\arxivbuild
\else
\IfFileExists{#1}%
{}%
{\GenericWarning{Marked file (#1) for inclusion in arxiv build does not exist}}
\fi%
}

\newcommand\markforarxiv[1]{%
\verifymarkedforarxivfile{#1}%
\write\arxivdeps{IncludeInArxiv: #1}%
}

\DeclareUrlCommand\UScore{\urlstyle{rm}}

\LetLtxMacro\oldincludegraphics\includegraphics
\renewcommand{\includegraphics}[2][]{%
\markforarxiv{#2}%
\oldincludegraphics[#1]{#2}}

\LetLtxMacro\oldincludepdf\includepdf
\renewcommand{\includepdf}[2][]{%
\markforarxiv{#2}%
\oldincludepdf[#1]{#2}}

\def\nfigure[#1,#2,#3]{
\begin{figure}
\vspace*{0mm}
\begin{center}

\includegraphics[width=\columnwidth]{#1} 
\caption[]{#2
} \label{#3}

\end{center}
\end{figure}}

\def\cfigure[#1,#2,#3]{
\begin{figure}
\vspace*{0mm}
\begin{center}

\includegraphics[width=3in]{#1} 
\caption[]{#2
} \label{#3}

\end{center}
\end{figure}}

\def\cfigurefour[#1,#2,#3]{
\begin{figure}
\vspace*{0mm}
\begin{center}

\includegraphics[width=4in]{#1} 
\vspace*{-3mm}\caption[]{#2
} \label{#3}
 
\vspace*{-5mm}
\end{center}
\end{figure}}

\def\cfiguretemp[#1,#2,#3]{
\begin{figure}
\vspace*{0mm}
\begin{center}

\includegraphics[width=3.5in]{#1} 
\vspace*{-3mm}\caption[]{#2
} \label{#3}
 
\vspace*{-5mm}
\end{center}
\vspace*{-2mm}
\end{figure}}

\def\wfigure[#1,#2,#3]{
\begin{figure*}
\vspace*{0mm}
\begin{center}
 \includegraphics[width=\textwidth]{#1} 
 \vspace*{-3mm}\caption[]{#2
} \label{#3}
 
\end{center}
\end{figure*}}

\def\threefigure[#1,#2,#3,#4,#5]{
\begin{figure*}
\vspace*{0mm}
\begin{center}

\begin{tabular}{ccc}
\includegraphics[width=2in]{#1} & \includegraphics[width=2in]{#2} &  \includegraphics[width=2in]{#3} \\
(a) & (b) & (c) \\
\end{tabular}
\vspace*{-3mm}\caption[]{#4
} \label{#5}

\vspace*{-5mm}
\end{center}
\vspace*{-2mm}
\end{figure*}}

\def\dcfigure[#1,#2,#3,#4,#5,#6]{
{
\begin{figure*}
\begin{center}
\begin{minipage}[c]{\columnwidth}{
\includegraphics[width=\columnwidth]{#1} 
\vspace*{0mm}\caption[]{#2} \label{#3} \
}\end{minipage}\hspace*{\columnsep}\
\begin{minipage}[c]{\columnwidth}{
\includegraphics[width=\columnwidth]{#4} 
\vspace*{0mm}\caption[]{#5}\label{#6} \
}\end{minipage}
\end{center}
\end{figure*}
}
}

\def\scfigure[#1,#2,#3]{
{
\begin{figure*}
\begin{center}
\begin{minipage}[c]{3.5in}{
\includegraphics[width=3.5in]{#1} 
}\end{minipage}
\caption[]{#2} \label{#3} \
\end{center}
\end{figure*}
}
}

\def\tableByTable[#1,#2,#3,#4,#5,#6]{
{
\begin{table*}
\begin{center}
\begin{minipage}[c]{3in}{
\centering
{#1}
\vspace*{0mm}\tabcaption[]{#2}\label{#3} \
}\end{minipage}\hspace*{\columnsep}\
\begin{minipage}[c]{3in}{
\centering
{#4}
\vspace*{0mm}\tabcaption[]{#5}\label{#6} \
}\end{minipage}
\end{center}
\end{table*}
}
}

\def\figureByTable[#1,#2,#3,#4,#5,#6]{
{
\begin{figure*}
\begin{center}
\begin{minipage}[c]{3in}{
\centering
\includegraphics[width=\textwidth]{#1}
\vspace*{0mm}\figcaption[]{#2} \label{#3} \
}\end{minipage}\hspace*{\columnsep}\
\begin{minipage}[c]{3.3in}{
\centering
{#4}
\vspace*{0mm}\tabcaption[]{#5}\label{#6} \
}\end{minipage}
\end{center}
\end{figure*}
}
}

\def\tableByFigure[#1,#2,#3,#4,#5,#6]{
{
\begin{figure*}
\begin{center}
\begin{minipage}[c]{4.3in}{
\centering
{#1}
\vspace*{0mm}\tabcaption[]{#2} \label{#3} \
}\end{minipage}\hspace*{\columnsep}\
\begin{minipage}[c]{2.2in}{
\centering
\includegraphics[width=\textwidth]{#4}
\vspace*{-0.35in}\caption[]{#5}\label{#6} \
}\end{minipage}
\end{center}
\end{figure*}
}
}

\def\doublecfigure[#1,#2,#3,#4]{
{
\begin{figure}
\begin{center}
\begin{minipage}[c]{1.5in}{
\begin{center}
\includegraphics[width=1.5in]{#1}
\end{center}
}\end{minipage}\hspace*{1em}\
\begin{minipage}[c]{1.5in}{
\begin{center}
\includegraphics[width=1.5in]{#2}
\end{center}
}\end{minipage}
\vspace*{0mm}\caption[]{#3} \label{#4} \
\end{center}
\end{figure}
}
}

\def\qcfigure[#1,#2,#3,#4,#5,#6]{
{
\begin{figure*}
\vspace*{0.2in}\
\begin{center}
\begin{minipage}[c]{3in}{
\includegraphics[width=3in]{#1} 
\vspace*{-3mm}
}
\end{minipage}\hspace*{0.5in}\
\begin{minipage}[c]{3in}{
\includegraphics[width=3in]{#2} 
\vspace*{-3mm}
}\end{minipage}

\begin{minipage}[c]{3in}{
\includegraphics[width=3in]{#3} 
\vspace*{-3mm}
}
\end{minipage}\hspace*{0.5in}\
\begin{minipage}[c]{3in}{
\includegraphics[width=3in]{#4} 
\vspace*{-3mm}
}\end{minipage}
\end{center}
\caption[]{#5}\label{#6}
\end{figure*}
}
}

\def\twfigure[#1,#2,#3,#4,#5]{
{
\begin{figure*}
\vspace*{0.2in}\
\begin{center}
\begin{minipage}[c]{6.5in}{
\includegraphics[width=6.5in]{#1} 
\vspace*{-3mm}
}
\end{minipage}

\begin{minipage}[c]{6.5in}{
\includegraphics[width=6.5in]{#2} 
\vspace*{-3mm}
}\end{minipage}

\begin{minipage}[c]{6.5in}{
\includegraphics[width=6.5in]{#3} 
\vspace*{-3mm}
}
\end{minipage}
\end{center}
\caption[]{#4}\label{#5}
\end{figure*}
}
}

\def\dwfigure[#1,#2,#3,#4]{
{
\begin{figure*}
\vspace*{0.2in}\
\begin{center}
\begin{minipage}[c]{6.5in}{
\includegraphics[width=6.5in]{#1} 
\vspace*{-3mm}
}
\end{minipage}

\begin{minipage}[c]{6.5in}{
\includegraphics[width=6.5in]{#2} 
\vspace*{-3mm}
}\end{minipage}

\end{center}
\caption[]{#3}\label{#4}
\end{figure*}
}
}

\def\dssfigure[#1,#2,#3,#4,#5,#6]{
{
\begin{figure*}
\vspace*{0.2in}\
\begin{center}
\begin{minipage}[c]{4in}{
\includegraphics[width=4in]{#1}
\vspace*{-3mm}\caption[]{#2} \label{#3} \
}\end{minipage}\hspace*{0.5in}\
\begin{minipage}[c]{2in}{
\includegraphics[width=2in]{#4}
\vspace*{-3mm}\caption[]{#5}\label{#6} \
}\end{minipage}
\end{center}
\vspace*{-0.4in}\
\end{figure*}
}
}

\def\dsfigure[#1,#2,#3,#4,#5,#6]{
{
\begin{figure*}
\vspace*{0.2in}\
\begin{center}
\begin{minipage}[c]{3in}{
\includegraphics[width=3in]{#1}
\vspace*{-3mm}\caption[]{#2} \label{#3} \
}\end{minipage}\hspace*{0.5in}\
\begin{minipage}[c]{3in}{
\hspace*{0.5in}\
\includegraphics[height=3in]{#4}
\vspace*{-3mm}\caption[]{#5}\label{#6} \
}\end{minipage}
\end{center}
\vspace*{-0.4in}\
\end{figure*}
}
}

\def\dsyfigure[#1,#2,#3,#4,#5,#6]{
{
\begin{figure*}
\vspace*{0.2in}\
\begin{center}
\begin{minipage}[c]{2.5in}{
\includegraphics[height=2.5in]{#1}
\vspace*{-3mm}\caption[]{#2} \label{#3} \
}\end{minipage}\hspace*{0.5in}\
\begin{minipage}[c]{2.5in}{
\includegraphics[height=2.5in]{#4}
\vspace*{-3mm}\caption[]{#5}\label{#6} \
}\end{minipage}
\end{center}
\vspace*{-0.4in}\
\end{figure*}
}
}

\def\dyfigure[#1,#2,#3,#4,#5,#6]{
{
\begin{figure*}
\vspace*{0.2in}\
\begin{center}
\begin{minipage}[c]{3in}{
\includegraphics[height=3in]{#1} 
\vspace*{-3mm}\caption[]{#2} \label{#3} \
}\end{minipage}\hspace*{0.5in}\
\begin{minipage}[c]{3in}{
\includegraphics[height=3in]{#4} 
\vspace*{-3mm}\caption[]{#5}\label{#6} \
}\end{minipage}
\end{center}
\vspace*{-0.4in}\
\end{figure*}
}
}

\def\dyoldfigure[#1,#2,#3,#4,#5,#6]{
{
\begin{figure*}
\vspace*{0.2in}\
\begin{center}
\begin{minipage}[c]{3in}{
\epsfysize=2.0in\
\hspace{0.5in}\
\epsfbox{#1}
\vspace*{-3mm}\caption[]{#2} \label{#3} \
}\end{minipage}\hspace*{0.25in}\
\begin{minipage}[c]{3in}{
\epsfysize=2.0in\
\hspace{0.5in}\
\epsfbox{#4}
\vspace*{-3mm}\caption[]{#5}\label{#6} \
}\end{minipage}
\end{center}
\vspace*{-0.4in}\
\end{figure*}
}
}

\def\cfiguredouble[#1,#2,#3,#4]{
\begin{figure}
\vspace*{0.2in}\
\begin{center}
\begin{minipage}[c]{1.5in}{
\epsfxsize=1.5in\
\epsfbox{#1}
}\end{minipage}\hspace*{0.1in}\
\begin{minipage}[c]{1.5in}{
\epsfxsize=1.5in\
\vspace{0.1in}\epsfbox{#2}
}\end{minipage}\vspace*{-0.10in} \caption[]{#3}\label{#4}
\end{center}
\vspace*{-0.4in}\
\end{figure}
}

\def\wpfigure[#1,#2,#3,#4]{
\begin{figure*}
\vspace*{4mm}
\begin{center}

\includegraphics[width=#4]{#1} 

\vspace*{-3mm}\caption[]{#2
} \label{#3}

\vspace*{-5mm}
\end{center}
\end{figure*}}

\def\wprfigure[#1,#2,#3,#4,#5]{
\begin{figure*}
\vspace*{4mm}
\begin{center}

\includegraphics[width=#4, angle=#5]{#1} 

\vspace*{-3mm}\caption[]{#2
} \label{#3}

\vspace*{-5mm}
\end{center}
\end{figure*}}

\def\DoubleFigureWSlide[#1,#2,#3,#4,#5,#6,#7,#8,#9]{
\begin{figure*}
\vspace*{#9}
\begin{center}
\begin{minipage}{#4}
\includegraphics[width=#4]{#1}
\vspace*{-3mm}\caption{#2
}\label{#3}
\end{minipage}
\hspace{2em}
\begin{minipage}{#8}
\includegraphics[width=#8]{#5}
\vspace*{-3mm}\caption{#6
}\label{#7}
\end{minipage}
\vspace*{-5mm}
\end{center}
\end{figure*}
}

\def\DoubleFigureW[#1,#2,#3,#4,#5,#6,#7,#8]{
\begin{figure*}
\vspace*{0in}
\begin{center}
\begin{minipage}{#4}
\includegraphics[width=#4]{#1}
\vspace*{-3mm}\caption{#2
}\label{#3}
\end{minipage}
\hspace{2em}
\begin{minipage}{#8}
\includegraphics[width=#8]{#5}
\vspace*{-3mm}\caption{#6
}\label{#7}
\end{minipage}
\vspace*{-5mm}
\end{center}
\end{figure*}
}

\def\DoubleFigureWHack[#1,#2,#3,#4,#5,#6,#7,#8]{
\begin{figure*}
\vspace*{0in}
\begin{center}
\begin{minipage}{3in}
\includegraphics[width=#4]{#1}
\vspace*{-3mm}\caption{#2
}\label{#3}
\end{minipage}
\hspace{2em}
\begin{minipage}{3in}
\includegraphics[width=#8]{#5}
\vspace*{-3mm}\caption{#6
}\label{#7}
\end{minipage}
\vspace*{-5mm}
\end{center}
\end{figure*}
}

\def\ddcfigure[#1,#2,#3,#4]{
\begin{figure*}
\begin{center}
\begin{minipage}[c]{\columnwidth}{
\includegraphics[width=\columnwidth]{#1} 
}\end{minipage}\hspace{0.5in}\
\begin{minipage}[c]{\columnwidth}{
\includegraphics[width=\columnwidth]{#2} 
}\end{minipage} \caption[]{#3}\label{#4}
\end{center}
\vspace{1pt}
\end{figure*}
}

\def\ddcfigureSlide[#1,#2,#3,#4,#5]{
\begin{figure*}
\vspace*{#5}\
\begin{center}
\begin{minipage}[c]{3in}{
\includegraphics[height=3in]{#1} 
}\end{minipage}\hspace{0.5in}\
\begin{minipage}[c]{3in}{
\includegraphics[height=3in]{#2} 
}\end{minipage}\vspace*{-0.10in} \caption[]{#3}\label{#4}
\end{center}
\vspace*{-0.4in}\
\end{figure*}
}

\def\cxfigure[#1,#2,#3]{
\begin{figure}
\vspace*{4mm}
\begin{center}
 
\epsfxsize=2.5in\
\epsfbox{#1}\
 
\vspace*{-0.10in}\caption[]{#2
} \label{#3}
 
\vspace*{-5mm}
\end{center}
\vspace*{-2mm}
\end{figure}}

\newif\ifremark
\long\def\remark#1{
        \begingroup%
        \dimen0=\columnwidth
        \advance\dimen0 by -1in%
        \setbox0=\hbox{\parbox[b]{\dimen0}{\protect\em #1}}
        \dimen1=\ht0\advance\dimen1 by 2pt%
        \dimen2=\dp0\advance\dimen2 by 2pt%
        \vskip 0.25pt%
        \hbox to \columnwidth{%
                \vrule height\dimen1 width 3pt depth\dimen2%
                \hss\copy0\hss%
                \vrule height\dimen1 width 3pt depth\dimen2%
        }%
        \endgroup%
}

\usepackage{xcolor}

\definecolor{cyanish}{rgb}{0,0.8,1.0}
\definecolor{orange}{rgb}{1.0,0.5,0.0}
\definecolor{pink}{rgb}{1.0,0.47,0.6}
\definecolor{light-gray}{gray}{0.95}
\definecolor{jiancolor}{RGB}{0,153,153}
\definecolor{mygreen}{RGB}{50,200,50}
\definecolor{pink}{rgb}{1.0,0.47,0.6}

\definecolor{commentgreen}{rgb}{0.0,0.5,0.0}

\newcommand{\boldparagraph}[1]{\vspace*{0.1ex}\noindent\textbf{#1}\hspace{1em}}

\newcommand{\etal}{\textit{et al.}}
\newcommand{\ignore}[1]{}

\newcommand{\reffig}[1]{Figure~\ref{#1}}
\newcommand{\reflst}[1]{Listing~\ref{#1}}
\newcommand{\refsec}[1]{Section~\ref{#1}}
\newcommand{\reftab}[1]{Table~\ref{#1}}
\newcommand{\reflns}[2]{Lines~\hyperref[#1]{\ref*{#1}-\ref*{#2}}}
\newcommand{\us}{\textmu{}s}
\newcommand{\x}[1]{$\times$}

\newif\ifcutforspace
\long\def\cutforspace#1{
\ifcutforspace%
        \begingroup%
        \dimen0=\columnwidth
        \advance\dimen0 by -1in%
        \setbox0=\hbox{\parbox[b]{\dimen0}{\protect{\em Cut For Space} #1}}
        \dimen1=\ht0\advance\dimen1 by 2pt%
        \dimen2=\dp0\advance\dimen2 by 2pt%
        \vskip 0.25pt%
        \hbox to \columnwidth{%
                \vrule height\dimen1 width 3pt depth\dimen2%
                \hss\copy0\hss%
                \vrule height\dimen1 width 3pt depth\dimen2%
        }%
        \endgroup%
\fi}

\newcommand{\nova}[1]{NOVA} 
\newcommand{\csym}[1]{\texttt{#1}}

\newcommand{\cfunc}[1]{\mbox{\csym{#1}\hspace{-0.1em}\csym{()}}}

\newcommand{\malloc}{\cfunc{malloc}}
\newcommand{\free}{\cfunc{free}}
\newcommand{\mmap}{\cfunc{mmap}\xspace}
\newcommand{\clwb}{\csym{clwb}}
\newcommand{\sfence}{\csym{sfence}}

\newcommand{\xname}{Puddles\xspace}
\newcommand{\puddled}{\texttt{Puddled}\xspace}
\newcommand{\libpuddles}{\texttt{Libpuddles}\xspace}
\newcommand{\libtx}{\texttt{Libtx}\xspace}

\newcommand{\CALM}{CALM\xspace}
\newcommand{\strike}[1]{}
\newcommand{\impr}[1]{}

\newcommand*\circledSolid[1]{\tikz[baseline=(char.base)]{
            \node[shape=circle,fill,inner sep=0.5pt] (char) {\textcolor{white}{#1}};}}

\newcommand{\statcalloverheadgetexistpuddle}{125.3~\us{}} %
\newcommand{\statcalloverheadgetnewpuddle}{1705.0~\us{}} %
\newcommand{\statcalloverheadping}{46.9~\us{}} %
\newcommand{\statcalloverheadreglogspace}{134.0~\us{}} %
\newcommand{\pmdktimetxnop}{142} %

\newcommand{\speedupformdemo}{4.7}
\newcommand{\maxspeedupformdemo}{10.1}

\newcommand{\speedupllpointerchase}{13.4}
\newcommand{\speedupbtreepointerchase}{3.1}

\newcommand{\maxspeedupycsb}{1.34}

\usepackage{xcolor}
\soulregister{\xname}{7}
\soulregister{\reftab}{7}
\soulregister{\reffig}{7}
\soulregister{\libpuddles}{7}
\soulregister{\puddled}{7}
\soulregister{\multicolumn}{7}
\soulregister{\texttt}{7}

\usepackage{todonotes}

\makeatletter
\def\ifempty#1{%
  \protected@edef\z{#1}\expandafter\ifblank\expandafter{\z}%
}
\makeatother

\renewcommand{\hl}[1]{#1}
\let\maketitlesup\maketitle
\usepackage{xpatch}
\xpatchcmd{\maketitlesup}{\@mkteasers}{}{}{}
\xpatchcmd{\maketitlesup}{\@mkabstract}{}{}{}

\newif\ifannotated
\annotatedfalse

\newcommand{\annotation}[1]{}

\ifannotated
\newcommand{\added}[1]{{\color{ForestGreen}#1}}
\else

\newcommand{\added}[1]{#1}
\fi

\begin{document}

\ifannotated
  {
    \setlength{\parskip}{0.7\smallskipamount}
    \begin{multicols}{2}
      [{\begin{center}\Huge{\textbf{Cover Letter}}\end{center}}]
      \setcounter{section}{0}
      \input{cover-letter}
    \end{multicols}
  }
\fi

\def\strut{} %
\title{Puddles: Application-Independent Recovery and Location-Independent Data for Persistent Memory}
\thanks{\textit{\small{} To appear in the Proceedings of EuroSys 2024, Athens, Greece.}}

\acmConference[EuroSys]{European Conference on Computer Systems}{2024}{Athens, Greece}

\author{Suyash Mahar}
\affiliation{\institution{UC San Diego}\country{USA}}
\author{Mingyao Shen}
\affiliation{\institution{UC San Diego}\country{USA}}
\author{TJ Smith}
\affiliation{\institution{UC San Diego}\country{USA}}
\author{Joseph Izraelevitz}
\affiliation{\institution{University of Colorado Boulder}\country{USA}}
\author{Steven Swanson}
\affiliation{\institution{UC San Diego}\country{USA}}
\date{}

\setlength\marginparwidth{40pt}

\soulregister{\added}{7}

\setcounter{section}{0}
\begin{abstract}
  \ignore{Persistent memory (PM) forces a rethink of how we store data at rest
    because it undermines essential assumptions about how persistent memory (PM)
    data is accessed.  Deserialization overhead of traditional file formats
    incur unacceptable overheads because the persistent media under the file
    system is now both reasonably fast and already a memory.}  In this paper, we
	argue that current work has failed to provide a comprehensive and maintainable in-memory representation for persistent memory.
	\ignore{and that we should reimagine how we store persistent memory data.} 
	PM data should be easily mappable into a process address space,
  shareable across processes, shippable between machines, consistent after a
  crash, and accessible to legacy code with fast, efficient pointers as first-class
  abstractions.

  While existing systems have provided niceties like \mmap{}-based load/store
  access, they have not been able to support all these necessary properties due
  to conflicting requirements.

  We propose \emph{Puddles}, a new persistent memory abstraction, to solve these
  problems.  \xname provide application-independent recovery after a power
  outage; they make recovery from a system failure a system-level property of
  the stored data rather than the responsibility of the programs that access it.
  Puddles use native pointers, so they are compatible with existing code.
  Finally, \xname implement support for sharing and shipping of PM data between
  processes and systems without expensive serialization and deserialization.

  \added{Compared to existing systems, \xname{} are at least as fast as and up
    to \maxspeedupycsb\x{} faster than PMDK while being competitive with other
    PM libraries across YCSB workloads. Moreover, to demonstrate \xname' ability
    to relocate data, we showcase a sensor network data-aggregation workload
    that results in a \speedupformdemo\x{} speedup over PMDK.}

\end{abstract}

\maketitle

\begin{spacing}{1}
\section{Introduction}
\label{sec:introduction}

Persistent Memory (PM) provides byte-addressability and large capacity, making
it ideal for memory-hungry applications like in-memory databases, graph
workloads, and big-data applications. Over the past decade, researchers have
proposed a host of systems that manage many of PM's idiosyncrasies and the
programming challenges it presents (\textit{e.g.}, persistent memory allocation
and crash recovery).

However, existing PM programming systems are built on a patchwork of
modifications on the memory-mapped file interface and thus make several
compromises in how persistent data is accessed. These systems use custom pointer
formats, handle logging through ad-hoc mechanisms, and implement recovery using
diverse \added{but incompatible} logging and transactional semantics.

\ignore{However, existing systems exhibit great diversity in how they format,
  manage, and organize PM data. While systems agree on the most fundamental
  access method---they all memory-map PM files to avoid serialization
  overhead---they diverge on nearly all other aspects of their design: how,
  when, and where PM is allocated, addressed, and recovered; what kind of logs
  are used and how to replay them (or roll them back); whether persistent data
  can move around in the address space; and how the pointers are implemented.}

In this paper, we show that the design of existing PM libraries results in PM
programming models that severely limit programming flexibility and introduce
additional unnatural constraints and performance problems. \ignore{ compared to traditional programming
  models.}

For example, opening multiple copies of a pool that resides at a fixed address
would result in address conflicts.
For another, using {non-native (\textit{i.e.}, ``smart''
or
``fat''
)} pointers avoids the need for fixed addresses but adds performance overhead to
common-case accesses, makes persistent data unreadable by non--PM-aware code,
leaves \strike{normal} software tools (\textit{e.g.}, debuggers) unable to
interpret that data, and locks-in the PM data to a particular PM
library. Further, current implementations of fat-pointers do not allow multiple
copies of PM data to map simultaneously unless the PM library first translates
all pointers in one of the pools\ignore{can translate all pointers in the
  pool}. Finally, enforcing crash consistency in the application requires that
after a crash, 1) the application is still available, 2) the application still
has write permissions for the data (even if the application only wants to read
it), and 3) the system knows which application was running at the time
  of the crash--none of which are true in general.

Today's PM programming libraries thus leave critical data integrity in the hands
of the programmer and system administrators rather than robustly ensuring those
properties at the system level.  Further, existing PM programming libraries
restrict basic operations like opening cloned copies of PM data simultaneously,
reading PM data without write access, or using legacy pointer-based tools to
access PM data.  A storage system with these characteristics represents
  a step back in safety and ensuring data integrity compared to the
  state-of-the-art persistent storage systems -- namely filesystems.

To solve these problems, we propose a new persistent memory programming library,
\textit{\xname}. \xname solve these problems while preserving the speed and
flexibility that the existing PM programming interface provides.
\xname provide the following properties:

\newcommand{\quicklist}{}

\quicklist{}(a)~\emph{Application-independent crash-recovery}: PM recovery after
a crash in \xname completes before \emph{any} application accesses the data.
Recovery succeeds even if the application writing data at the time of the crash
is absent after restart, no longer has the write permissions, or was just one of
the multiple applications updating the data at the time of the crash.

\quicklist{}(b)~\emph{Native pointers for PM data}: \xname use native pointers
and, thus, allow code written with other PM libraries or non-PM-aware code to
read and reason about it. Pointers are a fundamental and universal tool for
in-memory data structure construction. Changing their implementation for PM adds
runtime overhead of translation, requires specialized code to read PM data, and
stymies software engineering tools (\textit{e.g.}, compilers and debuggers do
not understand custom pointer formats used by PM libraries).

\quicklist{}(c)~\emph{Relocatability}: \xname can transparently relocate data to avoid any
address conflicts and thereby enable sharing and relocation of PM data between
machines. \ignore{This challenge is similar to generating position-independent,
  relocatable code in shared libraries.}

\xname is the first PM programming system that provides application-independent
recovery on a crash and supports both native pointers and relocatability while
providing a traditional transactional interface. \strike{Moreover, \xname do so
  while supporting a traditional PMDK-like transactional programming interface.}
Designing \xname, however, is challenging as native pointers, relocatability,
and mappable PM data are properties that are at odds with each other. \added{For
  example, native-pointers have traditionally prevented relocatable PM data, and
  non-mappable data like JSON does not support pointers.}

\ignore{(a)~Guaranteed consistent access after a crash, independent of the
  program that was running;

(b)~PM data accessibility for reading by non PM-aware code;

(c)~Location-independent data so it can reside anywhere in the address space;

(d)~be movable between systems  without serialization/deserialization.}

To resolve these conflicts, the Puddles system divides PM pools into
\emph{puddles}.  Each puddle is a small, modular region of persistent memory
(several MiBs) that the \xname library can map into an application's address
space. Puddles provide non-PM-aware applications access to PM data by allowing
programs to use native pointers. To support sharing puddles between processes
and shipping puddles between machines, puddles are relocatable---they can be
mapped to arbitrary virtual addresses to resolve address conflicts. \ignore{The
  relocation occurs when the puddle is mapped, and all pointers are
  automatically updated if needed.} To support relocation, puddles are
structured so that all pointers are easy to find and translate while, dividing
pools into puddles allows translation to occur incrementally and on
demand. The \xname library works in tandem with a privileged system
  service that allocates, manages, and protects the puddles.

To ensure that puddles are always consistent after a crash, puddle programs
register log regions with the system service and store the logs in those regions
in a format the service can safely apply after a crash. After a crash, the
system applies logs before \emph{any} application can access the PM data.
\xname' flexible log format can accommodate a wide range of logging styles
(undo, redo, and hybrid). While applications can access individual
  puddles, \xname supports composing them into seamless collections that
  resemble traditional PM pools, allowing applications to allocate data
  structures that span puddles.
\ignore{ and ensures that all mapped puddles in a \added{system
  (machine)} can coexist in a single, shared address space, allowing sharing
across processes, pointers between pools, and cross-pool transactions.}
\strike{Applications access puddles via a high-level library that builds on top
  of the puddle interface to provide familiar programming abstractions like
  failure-atomic transactions and logging.}

\ignore{Joe: this paragraph could just be cut to crisp the intro. It is really
 just supporting detail on the previous paragraph. Puddles combines three key
elements to provide these features. The first element is the puddle itself:
Puddles have a simple internal structure: the bulk
of the area is for data storage and a small header describes the location of
each pointer in the puddle, so they can be updated when it is (rarely) necessary
during puddle migration or re-location. All pointers in puddles are normal,
unadorned pointers. The second element is a privileged, user-space daemon that
allocates puddles and provides crash recovery. Application store logs in
designated log puddles using a flexible log format that the daemon can safely
apply after a crash on the application's behalf. Finally, the third element, or the
puddle libraries provide a low-level interface to the puddle daemon and a
higher-level PM programming library similar to PMDK.}

We compare \xname{} against PMDK and other PM programming libraries using
several workloads. \xname{} implementation is always as fast as and up to
\maxspeedupycsb\x{} faster than PMDK across the YCSB workloads. \xname{}' use of
native virtual pointers allows them to significantly outperform PMDK in
pointer-chasing benchmarks. \added{Against Romulus, a state-of-the-art
  persistent memory programming library that uses DRAM+PMEM, \xname, a PMEM-only
  programming library is between 36\% slower to being equally fast across the
  YCSB workloads.}

For linked-list traversal and B-tree search workloads, compared to PMDK,
\xname{} implementation is \speedupllpointerchase{}\x{} and
\speedupbtreepointerchase{}\x{} faster, respectively. \added{Moreover,
  support for relocatability allows \xname to perform data aggregation on copies
  of PM data without expensive serialization or reallocation, resulting in a
  \speedupformdemo\x{} speedup over PMDK.}

\section{Limitations of Current PM Systems}\label{sec:pm-programming-challenges}
\newcommand*{\boldcheckmark}{%
  \textpdfrender{
    TextRenderingMode=FillStroke,
    LineWidth=.5pt, %
  }{\checkmark}%
}
\newcommand*{\boldbigtimes}{%
  \textpdfrender{
    TextRenderingMode=FillStroke,
    LineWidth=.5pt, %
  }{\bigtimes}%
}

\newcommand{\TblBgGreen}{\cellcolor{ForestGreen!15}}
\newcommand{\TblBgRed}{\cellcolor{Maroon!15}}
\newcommand{\TxtGreen}{\color{ForestGreen}}
\newcommand{\TxtRed}{\color{Maroon}}
\newcommand{\supported}{\TblBgGreen\TxtGreen\boldcheckmark}
\newcommand{\unsupported}{\TblBgRed\color{Maroon}$\boldbigtimes$}

\newcommand*\rot[1]{\rotatebox[origin=c]{90}{#1}}
\newcommand{\tablefont}{\fontsize{7}{10}\selectfont}
\newcommand{\threecenter}[1]{{\rot{\parbox{1.3cm}{\setstretch{0.1}\tablefont\centering{}{#1}}}}}

\begin{table}[]%
\tablefont{}
\renewcommand{\arraystretch}{0.9}
\caption{\added{\textbf{Puddles vs. recent PM programming libraries.}}}
\label{tab:related-pm-libraries}
\centering
\setlength\tabcolsep{7pt} %
\begin{tabular}{|c|c|c|c|c|c|c|}\hline
                                                      &                                      &                               &                                                &                                      &                                      &                     \\[-1em]
                              \multirow{1}{*}{System} & \threecenter{Transactional                                                                                                                                                                                                \\Support} & \threecenter{Native Pointers} & \threecenter{Application Independent Recovery} & \threecenter{Object\\Relocatability} & \threecenter{Region\\Relocatability} & \threecenter{Cross-pool Transaction} \\
                                                      &                                      &                               &                                                &                                      &                                      &                     \\[-1.2em]\hline
\renewcommand{\arraystretch}{0.8}
PMDK~\cite{pmdk}                                      & \supported                           & \unsupported                  & \unsupported                                   & \unsupported                         & \supported                           & \unsupported        \\
TwizzlerOS~\cite{bittman2020twizzler}                 & \supported                           & \unsupported                  & \unsupported                                   & \supported                           & \supported                           & \unsupported        \\
Mnemosyne~\cite{mnemosyne}                            & \supported                           & \supported                    & \unsupported                                   & \unsupported                         & \unsupported                         & \supported          \\
NV-Heaps~\cite{nvheaps}                               & \supported                           & \unsupported                  & \unsupported                                   & \unsupported                         & \supported                           & \unsupported        \\
Corundum~\cite{corundum}                              & \supported                           & \unsupported                  & \unsupported                                   & \unsupported                         & \supported                           & \unsupported        \\
Atlas~\cite{atlas}                                    & \supported                           & \supported                    & \unsupported                                   & \unsupported                         & \supported                           & \unsupported                 \\
Clobber-NVM~\cite{clobbernvm}                         & \supported                           & \supported                    & \unsupported                                   & \unsupported                         & \supported                           & \unsupported        \\\hline
\textbf{\xname}                                       & \textbf{\supported}                  & \textbf{\supported}           & \textbf{\supported}                            & \textbf{\supported}                  & \textbf{\supported}                  & \textbf{\supported} \\\hline
\end{tabular}
\end{table}

Current persistent memory programs suffer from a host of problems that limit
their usability, reliability, and flexibility in ways that would be unthinkable
for more mature data storage abstractions. In particular, they rely on the
program running at the time of the crash for recovery, use proprietary pointers
that lock data into a single application or library, and place limits on the
combination of pools (i.e., files) an application can have open at one time.  

A novel file system with similar properties would garner little notice as a
serious storage mechanism, and we should hold PM systems to a similar standard.

\added{To understand the limitations and fragmented feature space of PM
  libraries, \reftab{tab:related-pm-libraries} compares several PM programming
  libraries across multiple axes. \xname is the only PM programming library that
  supports features like application-independent recovery, object relocatability
  (moving individual objects in a process's address space), region
  relocatability (moving groups of objects), and the ability to modify any
  global PM data in a transaction, features that users expect from a mature
  persistent storage programming system.

  The rest of the section examines problems that are endemic to existing PM
  programming solutions.

Next, in this section, we will look closely at
these problems that plague current PM programming solutions and understand how
they hold back PM applications.}

\subsection{PM Crash Recovery is Brittle and Unreliable}
When an application crashes, current PM programming libraries require the user
to restart the application that was running at the crash time to make its data
consistent. This design decision breaks the common understanding of data
recovery. 

For example, if a PDF editor crashes while editing a PDF file stored in a
conventional file system, the user can reopen the file with a different PDF
editor and continue their work. With current PM programming libraries, this is
not possible. The user must re-run the same program again, or the data is
inconsistent.

\added{This problem may seem benign, but this crash-consistency model relies on
  several assumptions that do not hold in general---like the availability of the
  original writer application and need for write access after a crash. The net
  result is an ad hoc approach to ensuring data consistency that is far removed
  from what state-of-the-art file systems provide. }

Indeed, ensuring recovery may not be possible at all in some circumstances.

For example, the user might lose write access to the data if their credentials
have expired, preventing them from opening the file to perform
recovery. Alternatively, the original application may no longer be available
either \added{because the licenses have expired,} OS and PM library updates have
changed the transactional semantics, \ignore{and cannot replay/undo the
  transactions} or if the file is restored from a backup on another system or
the physical storage media is moved to a new system. \ignore{Finally, the
  application may no longer be available after the crash.}  If any of these
assumptions fail, recovery will be impossible, and the data will be left in an
inconsistent state.

PMDK, the most widely
used PM library, illustrates how a lack of permissions can prevent recovery. In
PMDK, recovery is triggered only after the application restarts and reads the
same PM data; otherwise, the data is inconsistent. When the inconsistent data is
eventually read, PMDK looks for any incomplete transactions to recover the PM
data to a consistent state.  PMDK thus needs both read and write permissions to
the data before the application can read it again.

\subsection{PM Pointers are Restrictive and Inflexible}

\nfigure[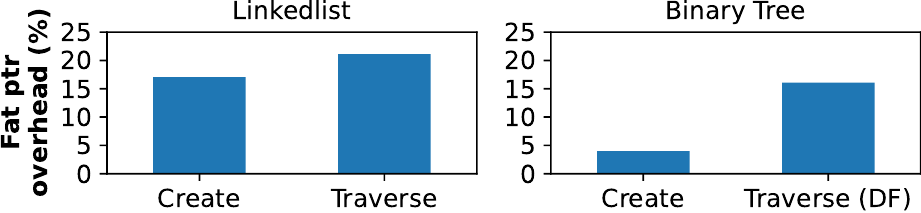,{Linkedlist and binary tree creation
  and traversal microbenchmarks, showing overhead of fat pointers vs.\ native
  pointers. Single-threaded workload. Linked list's length:
  $2^{16}$, and tree height: $16$},fig:fat-ptr-overhead] %

Persistent memory enables pointer-rich persistent data, but existing PM systems require programmers to
choose between non-optimal options: (a) use fat-pointers (base+offset) or
self-relative pointers and sacrifice performance on pointer dereference, or (b)
use native pointers and abandon relocatability.

Because fat pointers need to be translated to the native format on every
dereference, they suffer from a significant performance overhead. Further, the
large size of these pointers (in most cases, 128 bits) results in a worse cache
locality. \reffig{fig:fat-ptr-overhead} shows the overhead of fat pointers over
native pointers when creating and traversing a linked list and a binary tree. Fat
pointers show up to 16\% runtime overhead and result in an 18\%
higher L1 cache miss rate for the binary search tree microbenchmark.

Finally, using a non-native format for pointers makes them opaque and
uninterpretable to existing tools like compilers and debuggers.

\subsection{PM Data is Hard to Relocate and Clone}
Regardless of the pointer format choice, PM data is hard to
relocate. Consequently, with existing PM systems, users cannot create copies of
PM data and open them simultaneously, as the copies would either map to the same
address (with native pointers), or have the same UUID (with fat pointers).
Likewise, while some pointer schemes (e.g., self-relative~\cite{nvheaps}) allow
for relocation, they require relocating the entire pool at once and do not
support pointers between pools.

When using native pointers, cloned PM data contains conflicting pointers, and
the library has no way of rewriting them as the application does not know where
the pointers are. A similar problem exists with fat pointers: the application
would need to rewrite the base address of each pointer which is impossible in
current PM programming systems\ignore{ with no visibility into where the pointers in a
pool are}.

For example, the most widely used PM library, PMDK~\cite{pmdk}, identifies each
``pool'' of PM with a UUID and embeds that UUID in its fat pointers.  This
design requires a specialized tool to copy pools because the copy needs a new
UUID and all the pointers it contains need updating. PMDK thus prevents users
from opening multiple copies of a pool by checking if the UUID of the pool was
already registered when it was first opened. Further, the design also disallows
pointers between pools.

With persistent memory becoming more ubiquitous
with the emergence of CXL-based memory semantic SSDs~\cite{samsung-ssd} and
ReRAM-based SoCs~\cite{reram-soc}, beyond just Intel's Optane, 
the challenges of current persistent memory programming
remain present.  

\section{Overview}
\label{sec:overview}

The \emph{Puddles} library is a new persistent memory library to access PM data that
\hl{ supports application-independent recovery, and implements cheap,
  transparent relocatability, all while supporting native pointers.  To provide
  these features, \xname{} implement system-supported logging and recovery, a
  shared, machine-local PM address space for PM data, and transparent pointer
  rewriting to resolve address space conflicts.} In \xname, every application
that needs to access its data does so by mapping a puddle in its \added{virtual}
address space.

\subsection{Pools and Puddles}\label{sec:puddles-and-pools}
\added{Pools in the Puddle system are named collections of persistent memory and
  act as the programmer's interface to allocate and deallocate objects on PM,
  just like traditional PM pools. Pools automatically acquire new memory for
  object allocation and logging and free any unused memory to the system. 

  Pools are made of constituent puddles that are mappable units of persistent
  memory in the Puddle system.  While smaller than an average PM pool, puddles
  can span multiple system pages to accommodate large data structures. Moreover,
  despite puddles being non-resizable, a Pool as a collection of puddles permits
  the storage of multiple, often related data structures. Finally, pools enable
  PM data sharing across machines through a shareable, in-memory representation
  of their component puddles.}

\ignore{Puddles are mappable units of persistent memory data. A puddle is
  generally smaller than a typical persistent memory pool, but can be many huge
  pages large to accomodate large data-structures. Puddles cannot be dynamically
  resized, but, applications can combine multiple puddles into a \emph{pool} to
  store multiple, often related data-structures.  Pools automatically acquire
  new puddles for object allocation and logging and return free puddles to the
  system. Pools also enable applications to share PM data between machines by
  creating a shareable, in-memory representation of the constituent puddles.}

\subsection{\xname Implementation}
The Puddle system consists of three major system components
(\reffig{fig:puddles-arch}) that work together to provide application support
for mapping and managing puddles. \ignore{ that work together to provide the
  principles of \CALM. A privileged daemon (\puddled) provides system support,
  and two shared libraries that provide a simple programming interface on top of
  the daemon primitives.}

\nfigure[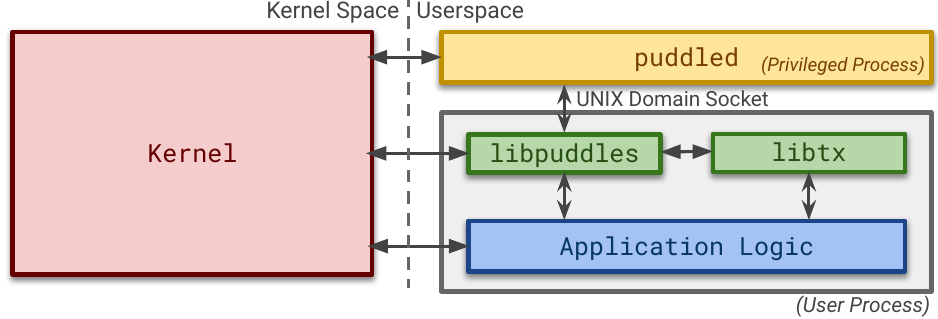,{The \xname system includes \puddled for system-supported
    persistence, \libpuddles, and \libtx for a simple programming interface on
    top of \puddled's primitives.},fig:puddles-arch]

\begin{enumerate} [leftmargin=18pt, rightmargin=0cm,itemsep=-1pt]
\item \emph{\puddled} is the privileged daemon process that manages access to
  all the puddles in a machine. \puddled implements access control and provides
  APIs for system-supported recovery and relocating persistent memory data.

\item \emph{\libpuddles} talks to \puddled and provides functions to allocate
  and manage puddles and pools. \ignore{Pools are self-contained with limited, and often, no
    incoming or outgoing pointers.}

\item \emph{\libtx} is a library that builds on the puddles semantics provided
  by \libpuddles. \libtx provides support for failure-atomic transactions that
  resemble the familiar PMDK transactions.
\end{enumerate}

Together, \libpuddles and \libtx provide a PMDK-like interface where the
application opens pools, allocates objects, and executes transactions without
managing or caring about individual puddles.

\ignore{These three parts work together to provide application support for
  mapping and managing puddles.}

\nfigure[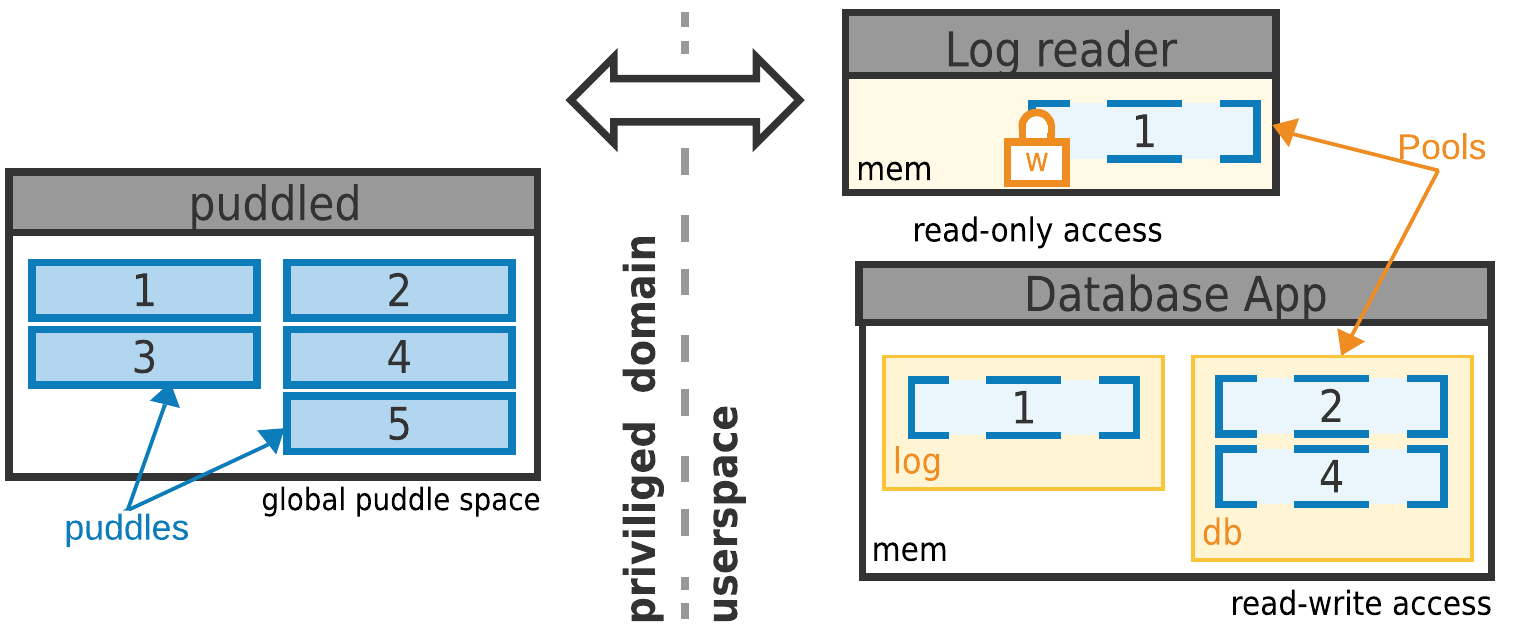,{\xname system overview. Each
  application talks to the \xname daemon (\puddled) to access the puddles in the
  system. Applications might map the same puddle with different
  permission.},fig:puddles-arch-overview]

\reffig{fig:puddles-arch-overview} shows an example database application that
demonstrates the benefits of \xname' approach where the database and logs are
partitioned into pools.  The application manages a PM database and writes event
logs using the \emph{Database app}. A separate \emph{Log reader} process has
read-only access to the event logs. Since both the database and the event logs
are part of the same global persistent space of a machine, the application can
write to both the database and the event log in the same transaction. The
application can also have pointers between the event log and the database, and
the \xname system would make sure that they work in any application with
permission to access the data.

\ignore{\xname support (1) \emph{Consistency} by allowing the applications to
  register recovery logs with the system and guaranteeing automatic recovery to
  a consistent state after a crash. (2) \emph{Accessibility} by using
  native-pointers unlike existing PM programming systems that implement fat- or
  offset-based pointers, \xname' data can be is accessible to existing volatile
  applications like compiler and debuggers. (3) \emph{Location independence} by
  reserving part of the address space for PM data and automatically translating
  pointers to move data structures to a different part of the reserved red
  address space. (4) \emph{Mappable}. Since all application data is stored
  directly in the in-memory format, libpuddles does not need any
  serialization/deserialization. (5) \emph{Movable}, since \xname are aware of
  all the in-memory pointers, \xname allow moving data-structures between
  address spaces (machines) and possibly relocates them to avoid address space
  conflicts in the new address space.

Next, we describe the details of each of these features:}

\subsection{Application Independent Recovery.}
In \xname, the application specifies how to recover from a failure, and the
system is responsible for recovering the data after a crash. Applications use
\xname{}' logging interface to register logging regions with \puddled.  The
logging interface is expressive enough to encode \added{undo, redo, and hybrid
  logging schemes}. 

In \xname, which component applies the logs depends on the context: During
normal execution, the application applies the
logs, %
but after a crash, the system applies them on the application's behalf.  In the
common case, the only additional overhead for the application is the one-time
cost of registering the logging region. This interface adds negligible overhead
to %
standard logging costs, similar to PMDK or other PM
programming libraries.

\ignore{Furthermore, \xname
  provides a shared memory interface to avoid expensive communication overhead
  during logging. Applications register a memory region as the logspace with the
  system and then write the logs directly to that memory region, similar to
  traditional persistent memory libraries.}

\subsection{The Puddle Address Space.}
\puddled maintains a machine-wide shared persistent memory space that all
puddles in a system are part of. At any time, an application only has parts of
the puddle address space mapped into its virtual address space. A single
persistent memory space \added{in a machine} allows \xname to support cross-pool
pointers and cross-pool transactions.

Applications allocate and request access to puddles from \puddled{}, which
grants them the ability to map the puddle into their virtual address
space. \ignore{While the \xname{} system allows applications to manage individual
  puddles, it also supports ``pool'' like API, where the application can perform
  \malloc{} seamlessly across multiple puddles.}

The puddle address space is divided into virtual memory pages where the puddles
are allocated as contiguous pages.  \hl{This global PM range only contains the
  application's persistent data; other parts of the application's address space,
  like the text, execution stack, and volatile heaps are still managed using the
  OS-allocated memory regions.}  In our implementation of \xname, we reserve
1~TiB of address space as the global puddle space \added{at a fixed virtual
  address, disregarding Linux's ASLR for the address range}. This range is
implementation-dependent and is limited only by the virtual memory layout.

\subsection{Native, Relocatable, and Discoverable Pointers.}
\label{subsec:pointers}
Puddles contain normal (\textit{i.e.},
neither smart nor fat) pointers to themselves or other puddles.  This ensures
that normal (non-PM-aware) code can dereference the pointers and read data
stored in puddles.

To ensure pointers are meaningful, each puddle must have a current (although not
fixed) address that is unique in the machine.  

This requirement raises the possibility of address conflicts: If an external
puddle (e.g., transferred from another machine) needs to be mapped, its current
address may conflict with another pre-existing puddle.  In this case,
\libpuddles{} will rewrite the pointers when mapping the new puddle into the
\added{application's} address space. To be able to rewrite pointers, \libpuddles
stores the type information with allocated objects, allowing it to quickly
locate all pointers to support on-demand, incremental relocation
(see \refsec{subsec:relocation}).

\ignore{Unlike traditional PM libraries that do not allow creating copies of PM
  data or exporting part of the data structure, \xname support the ability to
  relocate and export arbitrary part of the PM address space.  To handle
  accesses to unmapped puddles, \xname register a userspace fault handler for
  the entire unified PM space and transparently services them by fetching
  missing puddle from the OS.}

\ignore{While \xname{} has a single
unified global PM space across all the applications, an application typically
only has a few puddles mapped into its address space. \xname registers a
userspace fault handler and transparently services page faults for unmapped
puddles by getting them from the system.}

\subsection{\xname Programming Interface}\label{sec:programming-interface}

\added{To allocate objects, a pool provides traditional memory management-like API,
that is, PM analog of \malloc{} and \free{}. \hl{Allocations
  using this API can be serviced from any puddle in the pool with enough free
  space.} Pool's \malloc API takes as input the object's type in addition
to its size. To log PM data, the programmer can use either \texttt{TX\_ADD} for
an undo log or \texttt{TX\_REDO\_SET} for a redo log.}

\nfigure[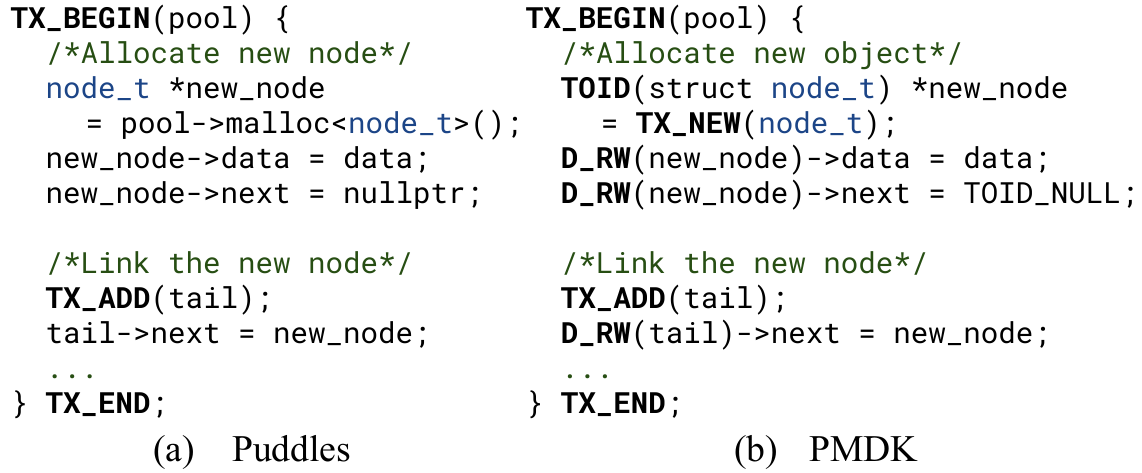,{List append example using (a) \xname,
  which uses virtual pointers, and (b) using PMDK, which uses base+offset
  pointers.},lst:puddles-pmdk-list-append]

\added{Transactions in \xname are similar to traditional PM transactions (e.g.,
  PMDK-like \texttt{TX\_BEGIN}...\texttt{TX\_END}, that mark the start and end
  of a transaction) and, thus, like other common PM transactions, do not provide
  support for concurrency nor IO in transactions and rely on the programmer to
  use mutexes. \reflst{lst:puddles-pmdk-list-append} is an example of a list
  append function written using both \xname and PMDK. The function code snippet
  allocates a new node on persistent memory and appends it to a linked
  list. \strike{\xname{}' use of virtual memory pointers reduces PM programming
    effort, requiring no extra calls to interpret the persistent pointers,
    unlike PMDK's \texttt{D\_RO} and \texttt{D\_RW}.} \xname{}' transactions are
  thread-local, but unlike PMDK, they support writing to any arbitrary PM data
  and are not limited to a single pool. }\strike{Nested transactions in \xname
  are flattened and committed only when the outermost transaction commits.}

\added{Finally, while \xname has a C-like API and is implemented using C++,
  similar to PMDK, \xname could be extended to support other managed languages
  like Java.}

\section{System Architecture}
Next, we discuss \xname' key capabilities that span across the \xname architecture,
with a focus on design decisions within our implementation.

\subsection{Crash Consistency}
\label{subsec:logging-and-crash-consistency}

\xname implement crash consistency by providing system support for crash
recovery, guaranteeing consistent data on PM access by any program \added{that
  correctly uses the \xname interface}. Further, \xname{}' centralized log
replay mechanism simplifies the recovery of shared PM data.

To guarantee consistent data on PM access after a crash, the system needs to be
able to replay the application's crash-consistency logs \added{provided the
  application correctly uses the logging interface.} Before an application
modifies any data, \xname library communicates the location and format of its
logs with the daemon for the daemon to use these logs during recovery. Further,
the logging format (a) needs to be able to support undo, redo, and hybrid
logging, (b) should be safe to execute independently of the application after a
crash, and (c) should not add significant runtime overhead.

\hl{To solve these challenges, \xname{} implement a novel, asynchronous, and
  performant logging format described next.}

\ignore{by offering expressive log entry format that supports
  per-log-entry undo and redo logging that registered only once with the
  system. To ensure security, \xname perform runtime and recovery security
  checks. }

\ignore{\xname' logging interface is expressive by supporting per log entry
  undo- or redo- mode, various logging techniques without adding significant
  overhead. This section describes how \xname{} achieve this and the details of
  \xname{}' system-supported logging interface and recovery mechanism.}

\cfigure[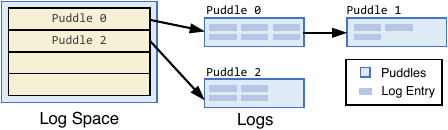,{Application registers a logspace with
  the system. A logspace space lists all puddles that the application uses to
  log data for crash consistency.},fig:logspace-and-logs]

\ignore{\paragraph{Logging and Recovery.}  In \xname, applications register the
  logs with \puddled, which applies them on the application's behalf after a
  crash and holds off any puddle access until all the logs are applied.
  Integrating crash recovery into the system guarantees that, after a crash,
  applications only read consistent PM data.}

\ignore{. advantages: First, the system guarantees that applications can only
  read consistent data after a crash. Second, the application does not need to
  be available, executed, or have write permissions to ensure that recovery
  occurs.  Finally, the applications sharing persistent memory do not have to
  coordinate recovery after a crash.}

\impr{Safety of log recovery}

\paragraph{Managing logs using log puddles and log spaces}
\xname organize logs using a directory, called a \emph{log space}, that tracks
all the active crash-consistency logs. Both the log space and the logs are
stored in designated puddles to simplify the implementation.
As shown in \reffig{fig:logspace-and-logs}, the
\emph{log space puddle} is a list of \emph{log space entries}, each identifying
a \emph{log puddle} that the application is using to store a log.  For instance,
an application might have one log puddle per thread to support concurrent
transactions. Each of these log puddles would have its own entry in the log
space.  Once registered, the application can update its log space or modify the
logs without notifying the daemon.

\ignore{At any point during execution, an application can have several active logs---each
of these logs are listed in the log space for \puddled to replay after a
crash. However, an application only needs one log space registered with
\puddled. This two-level list of logs allows the application to quickly allocate
a new log using an empty puddle without making an IPC request to \puddled. In
fact, the application only needs to communicate with \puddled once during its
execution in order to register a puddle as its log space. }

The puddle system allows the application to link multiple puddles to a log when
it runs out of space in its original puddle. \reffig{fig:logspace-and-logs}
shows an instance of this, where the first log in the log space spans two
puddles (Puddle 0 and 1).

\ignore{As explained in \refsec{sec:background}, PM programs use transactions to support
crash consistency and a log to record the information needed to return the data
to a consistent state after a crash. Typically, a PM application logs several 
values within a single transaction. Moreover, during execution, an application 
may run a large number of transactions, some which will also run in parallel. 
A straightforward approach to enabling system recoverability is for applications
to send log entries directly to \puddled, but the large number of log entries
required makes doing so prohibitively expensive. \xname mitigates the
communication overhead by using an asynchronous logging interface whereby the
application registers a logging region with \puddled, and then directly writes
all log entries to the region. The daemon then simply reads and executes the logs 
from the registered location during recovery.}

\paragraph{Flexible logging format}
Applications use a wide range of logging mechanisms (undo~\cite{atlas,
  george2020go}, redo~\cite{mnemosyne}, and hybrid~\cite{nvheaps, corundum,
  pmdk}), and \xname must be flexible enough to support as many as possible. To
allow this, \xname' logging format is expressive enough to cover a wide range of
logging schemes and structured enough for \puddled to apply them safely after a
crash to ensure that any data modified during the recovery was for an
application that had permission to it before the crash.

To be able to implement a variety of logging techniques, \xname' logging
interface allows the application to \added{(1) mark log entries to be of
  different types (e.g., an undo or redo entry). (2) Disable log entries by
  their type} so \puddled will skip them during recovery.  (3) Specify recovery
order, e.g., recover undo-log entries in reverse order. (4) And, verify that the
log entry is complete and uncorrupted.

A log in \xname is a sequence of log entries and includes the metadata to
control their recovery behavior. To provide a flexible logging interface, each
log entry in \xname contains the virtual address, checksum, flags field, log
data, and the data size. \xname use a combination of \emph{sequence number} (one
for each log entry) and a \emph{sequence range} (one for each log) to control
the recovery behavior. For every log, the log entries that have their sequence
number within the log's sequence range are valid, allowing \libpuddles (or the
application) to \added{selectively} (and atomically) enable and disable
\added{specific types} of log entries.

Finally, to specify the recovery order of the log entries, the log entry format
also supports specifying which entry to recover in which order. For example, the recovery
should always recover undo-log entries in the reverse order.

The log's metadata includes a pointer to find
the next free log entry, a pointer to the current tail entry, and the maximum
size of the log. \ignore{Similarly, each log entry contains the logged data and
  the associated metadata. To control the recovery behavior, the log includes a
  value range that specifies which entry is recovered.}

\reffig{fig:log-entry-format} illustrates \xname{}' log and log-entry
layout. The ``\texttt{Sequence Range}'' in the log and the ``\texttt{Seq}''
field in log-entry control recovery behavior by specifying which entries will be
used during recovery.  The ``\texttt{order}'' field specifies the order in which
log entries will be applied (forward for redo logging, backward for undo
logging). ``\texttt{Next log Ptr}'' and ``\texttt{Last Log Entry Ptr}'' track
log entry allocation. And, the checksum, like in PMDK, allows the recovery
code to identify and skip any entry that only partially persisted because of a
crash.

\cfigure[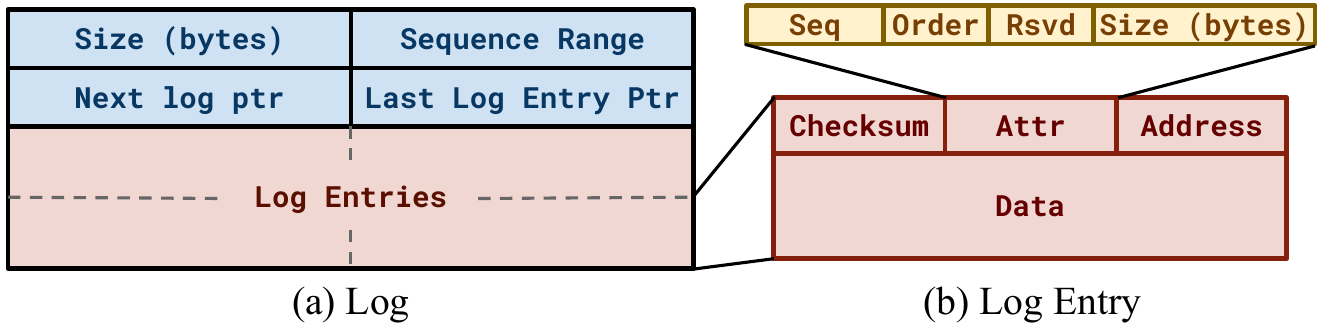,{\xname{}' log-entry and log format. \ignore{A log
  entry is replayed during recovery if an entry's \texttt{Seq} belongs to the log's
  sequence range.}},fig:log-entry-format]

\ignore{While \libpuddles{} and \libtx{} provide the interface for logging, similar to
traditional PM programming solutions, the application still needs to correctly
flush these logs to PM.}

\added{Finally, to keep transaction costs low, every thread caches the log
  puddle used on the first transaction of that thread and reuses it for future
  transactions. This prevents \libpuddles{} from allocating a new puddle and
  adding it to the log space on every transaction. Once the transaction commits,
  the log is dropped and is ignored by the \puddled.}

\paragraph{Example hybrid logging implementation}

To illustrate the flexibility of Puddle's log format, we will demonstrate how it
can implement a hybrid (undo+ redo) logging scheme. \added{Hybrid logging
  enables low programming complexity for application programmers that use undo
  logging while allowing libraries to implement their internals using faster but
  more complex redo logging. For example, PMDK uses hybrid logging to improve
  performance of allocation/free requests in
  transactions~\cite{pmdk-hybrid-logging}.}

\xname{}' expressive logging interface allows the application to implement
undo~\cite{atlas, george2020go}, redo~\cite{mnemosyne} or hybrid logging (which
uses both undo and redo logging simultaneously)~\cite{nvheaps, corundum, pmdk}
and control the recovery behavior.  \ignore{The two common logging techniques,
  the undo and the redo logging work in significantly different ways during the
  pre- and post-crash execution.} Undo logging requires the application to
back up the current memory value before modifying it. In comparison, redo logging
requires the application to write the new value directly to the log and apply
the log at the end of the transaction. \ignore{It is important to note here
  that} Undo-logged locations might have the updates durable in PM, while the
redo-logged locations are always unchanged until the changes are applied at
commit.

At runtime, \libpuddles allows the application to write undo- and redo-log
entries to a \puddled registered log. When the application calls transaction
commit, \libpuddles goes through all the log entries and processes them in
stages to be able to recover from a crash.

\begin{figure*}
  \vspace*{0mm}
    \def\subig[#1]{
      \includegraphics[width=0.9\linewidth]{hybrid-log-#1.pdf}\vspace{-0cm}
    }
    \newcolumntype{Y}{>{\centering\arraybackslash}X}

    \begin{center}
        \includegraphics[width=1\linewidth]{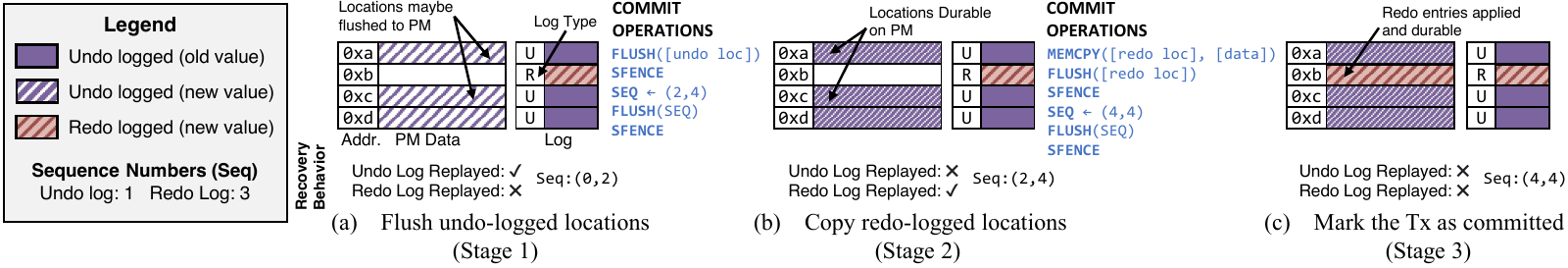}

    \caption[]{Three stages of hybrid logging TX commit and recovery. \added{Operations are instructions executed during commit stages.}}
    \label{fig:hybrid-log-desc}
    
  \end{center}
  \vspace*{-0.4cm}
\end{figure*}

\paragraph{Transaction commit.}
The transaction is committed in three stages by the userspace library, with no
involvement of \puddled. \added{This is shown in \reffig{fig:hybrid-log-desc}
   with the \sfence{} and \clwb{} ordering and the sequence numbers used
  for delineating the stages (elaborated on later in this section)}. The
first two stages work on the undo and redo logs, respectively, and the final
stage marks the log as invalid. These three stages are:

\begin{enumerate}[leftmargin=18pt, rightmargin=0cm,itemsep=-1pt]
\item \emph{Stage 1, Flush undo logged locations}
  (\reffig{fig:hybrid-log-desc}a). \libpuddles goes through the undo log entries
  and makes the corresponding locations durable on the PM.
\item \emph{Stage 2, Apply the redo log} (\reffig{fig:hybrid-log-desc}b). Once
  all the undo-logged locations are flushed to PM, \libpuddles starts copying
  new data from the redo logs. Redo logged locations were unchanged before the
  commit, so, \libpuddles copies the new data from the log entry to the
  corresponding memory location.
\item \emph{Stage 3, TX complete} (\reffig{fig:hybrid-log-desc}c). The
  transaction is complete, and all changes are durable. The log is marked as
  invalid.
\end{enumerate}

\paragraph{Recovery.}

Recovery is triggered on reboot after a dirty shutdown through the OS, where
\puddled applies a valid log from any incomplete PM transactions on behalf of the application.
The Puddles recovery process is three staged:

\begin{enumerate}[leftmargin=18pt, rightmargin=0cm,itemsep=-1pt]
\item \emph{Stage 1, Rollback}. First, \puddled applies all valid undo-log entries in reverse order. 

\hspace{0.3cm}In our example, at this stage, the undo log entries are not yet
  invalidated, however, some of the undo-logged locations might be
  durable. Thus, on recovery, \puddled can simply roll back the transaction by
  replaying the undo log. \puddled knows how to replay undo-logs by using the
  sequence range, sequence number, and recovery order.
\item \emph{Stage 2, Roll forward}. If the application crashed during stage 2, \puddled applies the redo-log entries from the log.

\hspace{0.3cm}In the example, all the undo-logged locations are
  durable, and \libpuddles might have applied some of the redo log entries. On a
  crash, the recovery would simply roll the transaction forward by resuming the
  redo log replay.
\item \emph{Stage 3, TX complete}. The TX was marked complete, and all changes
  are durable. No recovery is needed; any logs will be dropped.
\end{enumerate}

After a crash, the \added{daemon} compares each log entry's sequence number with
the log's sequence range to identify the active stage before the crash. In the
hybrid logging example, the application can assign sequence number 1 to the undo
log entries and 3 to the redo log entries (\reffig{fig:hybrid-log-desc}). This
assignment allows the application to define stages 1, 2, and 3 by setting the
log's sequence range to $(0,2)$ to only replay the undo logs, $(2,4)$ to only
replay the redo logs, and $(4,4)$ to replay neither.

\added{In \xname, regardless of whether an entry is an undo or redo log entry,
  to apply an active log entry, the daemon needs to only copy the entry's
  content to the corresponding memory location. For example, in undo logging, the
  entry contains the old data, copying its contents would ``undo'' the memory
  location. Similarily, copying contents of a redo log entry would apply the
  entry, resulting in a ``redo'' operation.}

\paragraph{Logging interface example.}

Although the application can directly write to the logs to be crash-consistent,
in \xname, programmers use PMDK-like transactions provided by \libtx to
atomically update PM data. To undo and redo log data within a transaction, the
programmer uses \texttt{TX\_ADD()} and \texttt{TX\_REDO\_SET()}, respectively.
Once the transaction commits, all changes are made durable.

\begin{figure}
  \includegraphics[width=\linewidth]{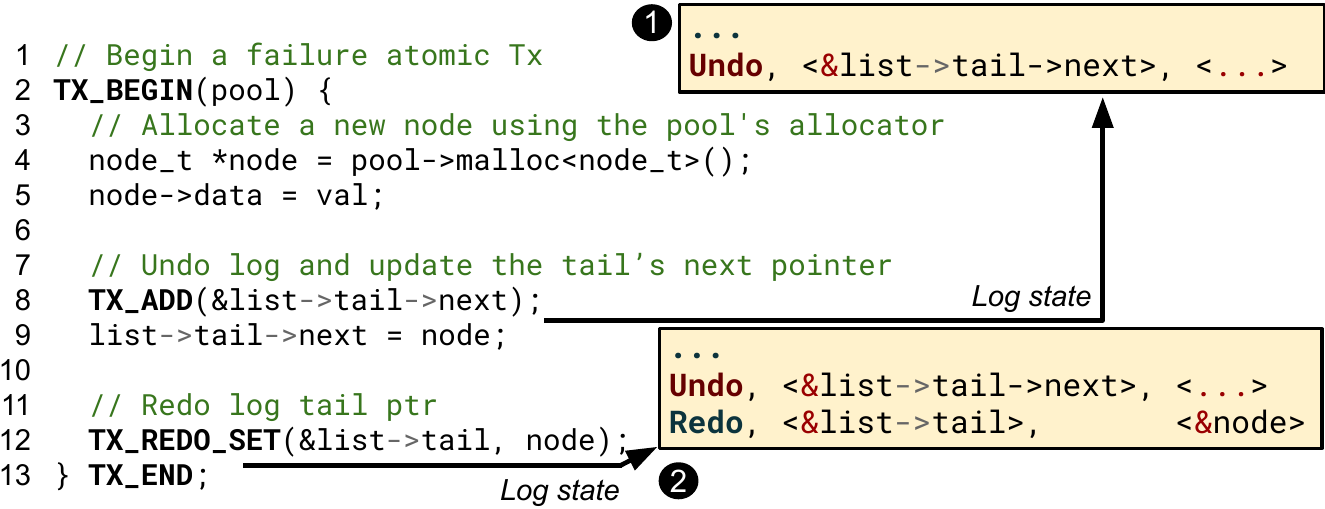}
  \caption{Linked List using \xname{}' programming interface along with the
    log's state after various operations. \ignore{\texttt{TX\_ADD()} undo logs a
      memory location, while \texttt{TX\_REDO\_SET()} creates a redo log
      entry.}}
  \label{fig:consistency-api}
\end{figure}

\reffig{fig:consistency-api} shows an example of a simple linked list
implementation to understand the programmer's view of the puddle logging
interface. The linked list implementation uses the puddle allocator to allocate
a new node (line 4). This new node is automatically undo-logged by the
allocator. Next, when the execution of line 8 completes, the log now contains a
new undo log entry for the next field of the current tail
(\circledSolid{1}). Next, the application redo logs the update to the list's
tail pointer (line 12). Being redo logged by the application, this update is
performed only on the log; the actual write location will be updated on the
transaction commit. Since the application uses hybrid logging, after line 12,
the log now contains both undo and redo log entries (\circledSolid{2}).

Once the execution reaches the \texttt{TX\_END}, \libpuddles executes the three
stages described in \reffig{fig:hybrid-log-desc} to commit the transaction and
make the changes durable. \ignore{\reffig{fig:consistency-api} lists the
  instructions executed by \libpuddles during each commit stage. To ensure
  correctness, \xname issues flush and fence instructions after modifying
  persistent memory locations.} \ignore{First, \libpuddles flushes all the undo
  logged locations to PM to make them durable and increments the log
  state. After this point, the undo log entries are no longer needed, and the
  log will only be redo-replayed in case of a failure. Next, \libpuddles copies
  all the updated values from the redo log to their corresponding locations on
  PM, flushes them, and drops the log. At this point, all changes are durable,
  and the transaction is committed. Thus, at any point in the execution, if the
  application crashes, it can be recovered to a consistent state.}

\paragraph{Logging design choices.}
An alternative (and superficially attractive) option to keeping a single log for
all puddles would be to keep per-puddle logs since this would make puddles more
self-contained.

Per-puddle logs, however, would have several problems.  First, concurrent
transactions on a puddle would require multiple logs per puddle, taking up
additional space and adding significant complexity in managing and coordinating these logs. Second, transactions that
span puddles (the common case in large data structures) would require a more
expensive multi-phase commit protocol.

\ignore{On the other hand, the only real limitation that our approach imposes is
  that the \puddled must recover puddles before exporting them, but since
  \puddled must be involved in the export anyway, this is not onerous.}

For logging, \xname' interface is limited to conventional per-location recovery
and does not support implementations that re-execute or resume
execution~\cite{clobbernvm, izraelevitz2016failure, liu2018ido}, semantic log
operations~\cite{pronto}, or shadow logging in DRAM and flushing it to
PM~\cite{liu2017dudetm, castro2018hardware, pangolin}. These systems use custom
logging techniques that require complex recovery conditions that make it
difficult to provide a unified interface. We intend to explore this support in
future work.

\annotation{We probably need to talk about how \xname deals with logs for freed
  puddles.}

\added{In addition to persistent memory locations, \xname logs can contain
  volatile memory locations that the applications apply on abort to keep
  volatile and persistent memories consistent with each other. During recovery
  after a crash, \puddled ignores these logs as the volatile state is lost.  }

\subsection{Location Independence}
\label{subsec:relocation}
\xname' ability to relocate PM data within the \added{virtual} address space
allows it to support location independence and movability with native
pointers. \xname thus allow applications to open multiple copies of PM data and
move data in the address space, properties expected from any storage system.

In the common case where the assigned address of PM data does not conflict with any existing puddles,
\libpuddles can simply map the puddle to the application's address
space. However, if the puddle's address (\refsec{subsec:pointers}) is already
occupied,
\xname support moving data in the global persistent address space.  \ignore{\xname
  accomplishes this by translating all the pointers in the PM data.} The
ability to move data on conflict is essential to support shipping PM data
between machines.

Pointer translation in \libpuddles works by incrementally rewriting pointers in
puddles. \libpuddles maps a puddle on demand and maintains a ``\emph{frontier}''
of puddles that are unmapped but have a reserved and available location in the
global persistent address space. Frontier puddles 1) are not yet mapped but
whose eventual location in the global persistent address space is reserved, and
2) are the target of a pointer in a mapped puddle that the mapped puddle points
to. As the application accesses data in the frontier puddles, \libpuddles
translate their pointers, maps them, and expands the frontier \added{to include
  puddles they contain pointers to.}

Specifically, when an application asks \libpuddles to map an unmapped puddle, \libpuddles maps
it to the puddle's assigned virtual address or, on a conflict, to an unreserved
range.  Next, \libpuddles iterates through all the pointers in the puddle and
checks if the pointer's destination address is already reserved. If the address
is reserved, \puddled assigns the puddle pointed by the pointer a new
address. This effectively relocates the target puddle in the \xname' global address
space, even if the puddle has not yet been opened or mapped to this
location. \hl{To accelerate finding pointers in a puddle's internal heap,
  puddles use allocator metadata to locate internal heap objects}
(\refsec{sec:allocator}).

Once all the pointers in a puddle are rewritten, \libpuddles makes it available
to the application to access.
At this point, only the puddle requested by
the application is mapped. If the application dereferences any pointer
\ignore{from the root object} that points to an unmapped puddle, it generates a
page fault. \libpuddles intercepts this page fault using the userspace page
fault handler (\texttt{userfaultfd}) and maps the faulting puddle to the
application's address space. By marking puddles that have been assigned a new
address and have not been mapped, \xname create a cascading effect of
\emph{on-demand} pointer rewrite where the pointers are only rewritten when the
data is mapped. Further, since all puddles in a machine are part of the same
virtual memory address range, \libpuddles can transparently catch access to any
unmapped data that is part of this range and map it to the application's address
space.

\added{Finally, \puddled persistently tracks puddles that were part of a
  frontier, including puddles that are not yet mapped. In case the machine
  crashes with some puddles unmapped, the next time one of the puddles from a
  frontier is mapped, the relocation process resumes.}

\hl{In summary, the \xname system relocates data on-demand within the
  \added{virtual} address space by mapping a new puddle and reserving space in
  the persistent address space for all the puddles pointed by the pointers in
  this puddle. Next, when these reserved but unmapped puddles are accessed,
  \libpuddles repeats these steps for the newly accessed puddle, thus creating a
  cascading effect.} 

\paragraph{Pointer maps}\label{par:pointer-maps}
For the puddle system to rewrite pointers, it needs to know their
location. \xname solve this problem by requiring the application to register
\emph{pointer maps} with \puddled for each persistent type used by the
application.  These pointer maps are simply a list, where each element contains
the offset of a pointer within the object and the type of the pointer.

To allow \xname to rewrite pointers, every allocation in \xname is associated
with a type ID, stored as a 64-bit identifier in the allocator's metadata
\hl{along with the allocated object}. Every class or struct with a unique name
corresponds to a unique type in \xname. Further, since allocations of a type
share their layout, \xname only need one pointer map per type. \added{To ensure
  each unique class's name results in a unique type, \xname rely on C++'s
  \texttt{typeid()} operator, just like PMDK~\cite{pmdk-typeid}. {typeid}s are
  generated using the Itanium ABI used by gcc and clang, which results in
  consistent typeid across at least gcc v8-12 and clang v7-12.}

The overhead of registering pointer maps with \puddled is negligible since the
number of unique objects an application uses is typically much greater than the
number of unique types it uses. Similar to the centralization of logs discussed
in \refsec{subsec:logging-and-crash-consistency}, we centralize the pointer maps
in \puddled to simplify puddle metadata management. \puddled stores the pointer
maps in a simple persistent memory hashmap along with its other metadata. While
pointer maps could be stored in each puddle for the types in the puddle, doing
so would require dynamic memory management of the puddle's metadata. Since the
overhead of storing the pointer map information with \puddled is low, and it is
easy for \puddled to export its pointer maps along with exported puddles, we
found the complexity of storing pointer maps in puddles rendered it not worth
pursuing.

\paragraph{Relocation on import.}
\xname allow sharing of PM data  by
``exporting'' part of the global persistent space. Once exported, PM data 
retains its in-memory representation, allowing \xname to ``import'' it back into
the same address space as a copy, or into a different machine.

\ignore{When an application imports new PM data into its address space and it
  conflicts with a reserved or an already allocated range (in case the data is
  re-imported as a copy), \xname automatically rewrite all the pointers to
  relocate it to a new address range. }When importing data, the application asks
\libpuddles to map a pool into its address space. \ignore{A pool contains a root
  puddle that acts as an entry to all the data in it. }  Pools are a collection
of puddles with a designated root puddle, the puddle that holds the pool's root
object. \xname support relocation on import by first mapping the root puddle of
the pool. Once \libpuddles maps the root puddle, it can begin its pointer
rewrite operation and relocate any conflicting data. 

\ignore{In \xname, when an application tries to read two copies of PM data,
  \xname automatically rewrites all the pointers allowing the operation to
  continue.}

Location independence in \xname extends to support movability by allowing the
application to export the underlying data in its in-memory form. Exporting pools
in \xname does not require any serialization and exports the raw in-memory data
structures. Once exported, the PM data can be re-imported into the machine's
global PM space with no user intervention. Existing PM programming solutions do
not allow applications to relocate PM data between pools or create copies of PM
data without reallocating and rebuilding all the contained data structures in a
new PM pool.

\ignore{Combined with object type IDs managed by the allocator, the pointer maps
  allow \puddled to enumerate all the objects in a heap and rewrites its
  pointers on an address conflict. Together, these operations allow \xname to
  seamlessly relocate PM data, thereby enabling the user to create multiple
  copies of the underlying PM data without concern for data placement.}

\ignore{\paragraph{Type information in \xname} A puddle stores the type
  information in the allocator metadata to track the allocation's type and uses
  it to translate pointers on an address conflict.}

\ignore{Now that we have looked at how \xname offer its properties to make PM
  programming more flexible, we will discuss the implementation in more detail.}

\paragraph{Referential Integrity}
\added{While the referential integrity of exported data is a concern, applications
  are expected to only export self-contained pools. In \xname, this can be
  accomplished by limiting inter-pool links or using programming language
  support to prevent inter-pool pointers.  Enforcing strong referential
  integrity guarantees within \xname itself is left as future work.}

\subsection{Puddle Implementation}\label{subsec:puddle-implementation}
\added{In the \xname system, puddles are contiguous regions of persistent memory
  that have a heap to store the application data and an associated header to
  store the allocator metadata.}  All puddles in a machine are part of the same
global PM address range. When an application starts, \libpuddles reserves this
address range in the application's virtual address space. Being regions of
memory, Puddles can be of any size in multiples of an OS page, but they cannot
grow or shrink once they are created.

Every puddle in the global puddle PM space has a 128-bit universally unique
identifier (UUID). A puddle has two parts, a header, and a heap. The header
stores the puddle's metadata information like the puddle's UUID, its size, and
allocation metadata. The heap is managed by the \libpuddles' allocator and
contains all allocated objects and their associated type IDs. We have configured
\xname to have 4~KiB of header space for every 2~MiB of heap space (0.2\%
overhead).

We implement puddle management in \puddled by leveraging the underlying
filesystem and avoid re-implementing a puddle-scale allocator.  For each puddle,
\puddled creates a file in the filesystem, accessible only by \puddled
itself. The filesystem, however, must support DAX (Direct Access) to directly
map the puddle contents into the virtual memory space. \ignore{Using the
  filesystem allows \puddled to offload the management of persistent memory
  space to the filesystem, making implementation substantially easier and less
  bug-prone. \puddled could instead manage the persistent memory space itself
  with some kind of puddle-scale allocator, but the filesystem already handles
  this task well, so we chose not to re-implement it.} If a request for a puddle
is permitted, \puddled returns a file descriptor for the puddle to the
application using the \texttt{sendmsg()} system call.

Had we implemented \puddled inside the kernel rather than as a privileged
daemon, we could have allowed \puddled to directly update the application's page
tables to map the puddle. This would reduce the overhead of sending the file
descriptor through the domain socket. However, we decided to leave \puddled in
user space because it makes it much easier for users to adopt \xname. Instead of
needing to install a custom-built kernel, users of \xname only need to run
\puddled and link to our libraries. \ignore{Likewise, while it is possible to
  allow applications to directly access the underlying file for a puddle, doing
  so would prevent \puddled from guaranteeing that all pointers are translated
  when a puddle is requested by an application.}

\subsection{Pools}\label{subsec:pools}
The \xname system provides a convenient \emph{pool} abstraction on top of
puddles to create data-structures that span puddles.  \ignore{This abstraction,
  or, a \emph{pool} is a collection of semantically related puddles usually
  limited to a single application.} \hl{Programmers use a pool's \malloc{}-like
  API to avoid needing to manually manage objects between puddles. \ignore{For
  example, a database application may store the database table and the metadata
  in separate pools.}

Internally, \puddled{} and \libpuddles{} identify a pool as a collection of
puddles and a designated ``root'' puddle. The root puddle of a pool is the
puddle that contains the root of the data structure contained in that pool.
\ignore{For example, a linked list's pool would have the head node of the list
  stored in the root puddle. Applications use the pool's root UUID to request
  access to the pool.}}

After \puddled verifies that the application has access permission to a pool,
the library receives the pool's root puddle and maps the puddle to its virtual
memory address space. \libpuddles then maps the puddle lazily using the
on-demand mapping mechanism described in \refsec{subsec:relocation}.

\added{Segmenting the persistent memory address space into small puddles to
  provide the pool interface enables Puddles to relocate, share, and recover
  individual objects with fine granularity, resulting in low performance and
  space overhead. For example, puddles limit the cost of pointer rewrite when
  importing large PM data, limiting the overhead to a few puddle at a
  time. Further, programmers are expected to maintain no incoming or outgoing
  pointers from a pool, making pools self-contained for export to a different
  machine. }
\annotation{SM: Moved this paragraph to the end of the subsection}

\subsection{Object Allocator}\label{sec:allocator}

While each puddle can be independently used to allocate objects, applications
typically use pools to allocate objects.  Using a pool makes it easier for the
application to package and send its data structures to a different address
space. Further, to track the allocation's type, pool's \malloc API takes as
input the type of the object in addition to its size.

Since the object allocator always allocates the first object at a fixed offset
(root offset) in the puddle, when the application asks \libpuddles for the root
object, \libpuddles can return its address using a simple base and offset
calculation.

Object allocations in puddles are handled in the userspace by \libpuddles,
similar to PMDK. \xname use a two-level allocator where per-type slab allocators
manage small allocations (< 256 B). Large allocations \ignore{and slabs} are
allocated from a per-puddle buddy allocator. Two-level allocator hierarchy
allows \xname to perform fast allocations of both large and small sizes.

\subsection{Access Control}\label{sec:access-control}

\added{In the puddle system, applications must not access puddles that they do
  not have permission to, while allowing \puddled to manage all puddles in a
  machine. To achieve this, \puddled stores each puddle in a separate file on
  the PM file system. These files are exclusively owned by \puddled, and no
  other process can access them. For applications to access a puddle, \puddled
  maintains a separate, application-facing, UNIX-like permission model.

  When an application requests access to a puddle over the UNIX-domain socket,
  puddled verifies the caller's access using its group ID and user ID. If
  approved, puddled returns a file descriptor for the requested puddle using the
  \texttt{sendmsg(2)} system call. This file descriptor serves as a capability,
  letting the application access the underlying puddle without any direct access
  to the underlying file. Upon receiving the file descriptor, \libpuddles maps
  the puddle to the application's address space and closes the file descriptor.

  While sending file descriptors simplifies puddle management and mapping, an
  application can still forward them to other processes. However, this
  limitation is inherent to the UNIX design, i.e., the same vulnerability applies
  to files, and thus, we assume a similar adversary model.

  Finally, as applications must communicate with \puddled to request access to
  puddles, \puddled starts before any other process in the system and controls
  access to PM data.}
\ignore{This check is similar to the permission checks done for a
  file by a filesystem. \xname use the Unix permission model with separate
  permissions for the owner, the group, and others. The current implementation
  supports read and write permissions.}  \ignore{In response to a puddle
  request, \puddled checks its own stored permissions for the puddle (these are
  different from the file permissions managed by the filesystem).}

\ignore{Using
  file descriptor restricts application's access to the particular puddle sent
  by}

\ignore{When requesting access to a puddle, if the user does not have both read and
write access, \puddled will return a permission error. However, in case the user
only has read access, and requests read on a puddle, \puddled will return a
read-only file-descriptor that the application would only be able to map without
the write access. In any other case, the application can map the region normally
and perform both read and write operations.}

\paragraph{Recovery}\label{sec:access-control-recovery}
\added{\puddled extends the puddle access control to recovery and prevents a
  process from using recovery logs to modify unwritable addresses. During log
  replay, \puddled recreates the mapping for the crashed process by mapping all
  puddles in the machine-local persistent address space. Recreating the puddle
  mapping limits \puddled's recovery to locations that the process had write
  permission to before the crash.} If \puddled identifies an invalid log,
instead of dropping the log, it will be marked as invalid and will not be
replayed as the PM data is possibly in a corrupted state. \added{While this may
  result in denial of service by a malicious application, the effect would be
  limited to the data accessible by the application.}

\added{Let us explore a scenario where a potentially malicious application logs
  data and then frees the corresponding puddle. In this context, Puddles ensures
  data integrity by only allowing access to applications with proper
  permissions. Two scenarios can arise: (1) a new application acquires the freed
  puddle but allows the original application access to the puddle, potentially
  risking data corruption during recovery. However, the malicious application
  already had access to the data before the crash, and thus the security
  guarantees are unaffected. (2) If the puddle is unallocated during recovery,
  or if another application acquires the puddle but does not permit the original
  application access to its data, the recovery after a crash will fail as the
  malicious application no longer has access to the puddle, preserving data
  integrity.}

\section{Results}
\label{sec:results}

\begin{table} %
  \fontsize{8}{10}
  \selectfont
  \caption{System Configuration}
  \label{tab:sysconfig}
  \centering
  {
    \setlength{\tabcolsep}{0.4em}
    \begin{tabular}{l|lll|l}
      \cline{1-2}\cline{4-5}
      \textbf{CPU \& HW Thr.} & Intel Xeon 6230 \& 20 && \textbf{Linux Kernel} & v5.4.0-89  \\
      \textbf{DRAM / PM}      & 93~GiB / 6\x{}128~GiB && \textbf{Build system} & gcc 10.3.0 \\\cline{1-2}\cline{4-5}
    \end{tabular}
  }
\end{table}

\xname perform as fast or faster than PMDK and are competitive with
state-of-the-art PM libraries across all workloads while providing
system-supported recovery, simplified global PM space, and relocatability. We
evaluate \xname using BTree, KV Store using the YCSB benchmark suite, Linked
List, and several microbenchmarks.

\ignore{All evaluations were performed on a single socket Intel Cascade Lake
  Processor with Optane DC-PMM.} \reftab{tab:sysconfig} lists the system
configuration. For all experiments, we use Optane DC-PMM in App Direct Mode.
All workloads use undo-logging for both the application and allocator data
logging except for the linked list workload.

\subsection{Microbenchmarks}\label{subsec:microbenchmarks}
This section presents three different measurements to provide insights into how
\xname perform: (a) performance of \xname{}' API primitives to compare and
contrast them with PMDK, (b) average latency and frequency of \puddled
operations, (c) and, the time taken by different parts of \xname{}'
relocatability interface.

\def\ReplaceUs#1{%
  \IfSubStr{#1}{~\us{}}{%
    \saveexpandmode\noexpandarg\StrSubstitute{#1}{~\us{}}{}
      \restoreexpandmode
  }{#1}
}
\def\ReplaceNs#1{%
  \IfSubStr{#1}{~ns}{%
    \saveexpandmode\noexpandarg
    \StrSubstitute{#1}{~ns}{}
    \restoreexpandmode
  }{#1}
}

\def\FmtLatUs#1#2#3{%
  {\saveexpandmode\noexpandarg\ReplaceUs{#1}/\ReplaceUs{#2}/#3
\restoreexpandmode}
}
\def\FmtLatNs#1#2#3{%
  {\saveexpandmode\noexpandarg\ReplaceNs{#1}/\ReplaceNs{#2}/#3
\restoreexpandmode}
}

\newcommand{\datasizes}{\texttt{8B/4kB}}

\begin{table}
  \fontsize{8}{10}
  \selectfont
 \caption{Mean latency of \xname and PMDK primitives.}
 \label{tab:avg-lat}
 \centering
 \begin{tabular}[t]{|l|c|c|}
   \hline
   \multicolumn{1}{|c|}{\multirow{1}{*}{\textbf{Operation}}} & \textbf{\xname}                & \textbf{PMDK}                    \\\hline\hline
   \texttt{TX NOP}                                           & 11.0~ns                        & \pmdktimetxnop{}~ns              \\\hline
   \texttt{TX\_ADD} (\datasizes)                             & 0.04/1.1~\us{}                 & 0.3/2.2{}~\us{}                  \\\hline
   \texttt{malloc} (\datasizes)                              & 0.1/6.8~\us{}                  & 0.4/0.4~\us{}                    \\\hline 
   \texttt{malloc}+\texttt{free} (\datasizes)                & \multirow{1}{*}{5.6/6.0~\us{}} & \multirow{1}{*}{2.0/3.0{}~\us{}} \\\hline
 \end{tabular}
\end{table}

\boldparagraph{API primitives.}\label{par:api-primitives} Measurements in
\reftab{tab:avg-lat} show that across most API operations, \xname outperform
PMDK.  To measure the overhead of starting and committing a transaction, we
measure the latency of executing an empty transaction -- \texttt{TX NOP}. Since
\xname' transactions are thread-local and do not allocate a log at the beginning
of a transaction, they are extremely lightweight. For an empty transaction,
\xname{}' overhead only includes a single function call to execute an empty
function.

For undo-logging operations (\texttt{TX\_ADD}), \xname have latencies similar to
PMDK. However, we observe slower allocations (\malloc{}) and de-allocations
(\free{}) for \xname. \added{The performance difference is an artifact of the
  implementation. For example, \xname uses undo logging while PMDK uses redo
  logging for the allocator.}

\boldparagraph{Daemon primitives.} Since the \xname system offers
application-independent recovery, it needs to talk to \puddled to allocate
puddles and perform other housekeeping
operations. \ignore{\reftab{tab:daemon-funcs} lists their average overhead.} The
daemon communicates with the application using a UNIX domain socket. On average,
a round-trip message (no-op) between the daemon and the application takes
\statcalloverheadping{}. Most daemon operations take in the order of a few
hundred microseconds to complete. \ignore{For example, allocating a new puddle
  takes on average \statcalloverheadgetnewpuddle{} and getting a reference to an
  existing puddle takes \statcalloverheadgetexistpuddle{}. These latencies
  includes the cost of updating the daemon's persistent state.}  

During execution, the function \texttt{RegLogSpace} is called once to register a
puddle as the log space and takes on average
\statcalloverheadreglogspace{}. \texttt{GetNewPuddle} and
\texttt{GetExistPuddle} are called every time the application needs a
puddle. Internally, \puddled manages each puddle as a file and returns a file
descriptor for puddle requests. Allocating a new file slows
down \texttt{GetNewPuddle}, and it takes considerably longer
(\statcalloverheadgetnewpuddle{}) than calls to \texttt{GetExistPuddle}
(\statcalloverheadgetexistpuddle{}). Even though the call to
\texttt{GetNewPuddle} is relatively expensive, \libpuddles mitigates their
overhead by caching a few puddles when the application starts.  Caching puddles
in the application avoids calls to the daemon when the application runs out of
space in a puddle. As we will see with the workload performance, even with
relatively expensive daemon calls, \xname outperform PMDK.

\added{Finally, in addition to the runtime overheads, recovery from a crashed
  transaction takes 110.1~\us{} in \xname.}

\boldparagraph{Relocatability primitives.} \ignore{\reftab{tab:daemon-funcs}
  also lists the overhead of \xname{}' primitives to support relocatability.} On
a request to export a pool, \puddled creates copies of the puddles and the
associated metadata (e.g., pointer maps). Data export cost, therefore, scales
linearly with the size of the PM data and includes a constant overhead per
puddle. Exporting a pool takes 0.3~s for 16~B and 0.5~s for 16~MiB of PM data in
our implementation. Importing data, on the other hand, is nearly free, as it
only includes registering the imported puddles with the daemon (1.5~ms for both
16~B and 16~MiB). After import, if the imported data conflicts with an existing
range, the puddle system automatically rewrites all the pointers in the mapped
puddle. During pointer rewrite, every pointer in the pool must be visited, so
runtime scales linearly with the number of pointers in the pool. \hl{Rewriting
  pointers takes 0.2~ms for 20 pointers, 1.6~ms for 2000 pointers, and 0.5s for
  2~million pointers.}

\added{\boldparagraph{Correctness Check.} To ensure the correctness of \xname'
  logging implementation, we inject crashes into \xname' runtime and run
  system-supported recovery. We do this for undo and redo logging and find that
  \xname recover application data to a consistent and correct state every time.}

\subsection{Workload evaluation}\label{subsec:workload-evaluation}

\nfigure[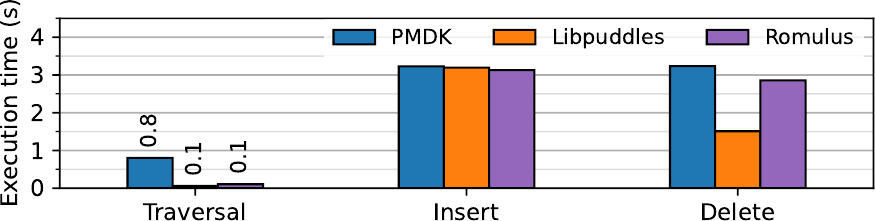,{\xname{}' performance against PMDK and Romulus
  for singly linked list (lower is better). Native pointers offer a significant
  performance advantage for \xname.},fig:linkedlist-perf]

\newcommand{\simplepara}[1]{\boldparagraph{#1}}

\added{To evaluate \xname's performance, this section includes results for
  several workloads implemented with \xname, PMDK, Romulus, go-pmem, and
  Atlas. Further, to understand the overhead of fat-pointers in PMDK, we used
  stack samples from PMDK workloads and find that the overhead of fat-pointers
  ranges from 8.5\% for btree, which has multiple pointer dependencies, to 0.76\%
  for the KV-store benchmark that uses fewer pointers per request by making
  extensive use of hash map and vectors. Finally, across workloads, the daemon
  primitives result in an additional overhead of about 0.2~ms. This overhead is
  primarily from registering the first log puddle during the transaction of the
  benchmark.}

\simplepara{Linked List} We compare \xname{}' implementation of a singly linked
list against PMDK \added{and Romulus}. \reffig{fig:linkedlist-perf} compares the
performance of three different operations (each performed 10 million times): (a)
Insert a new tail node, (b) delete the tail node, and (c) sum up the value of
each node. For the insert, all programming libraries have similar performance
with the exception of Romulus, but delete and sum in \xname outperform PMDK by a
significant margin. This performance gap is from the native pointers' lower
performance overhead and better cache locality in \xname{}. \added{In addition
  to \xname' undo logging implementation presented here, we evaluated a hybrid
  log implementation using undo logging for the allocator and redo logging for
  the application data and found the performance to be similar to the
  undo-logging only version, that is, within 5\%.}

\nfigure[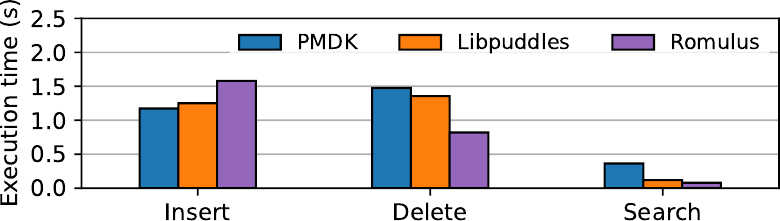,{Performance of \xname, PMDK, and Romulus's
  implementation of an order 8 Btree (lower is
  better).},fig:btree-perf]

\simplepara{B-Tree} \reffig{fig:btree-perf} shows the performance of an
identical order 8 B-Tree implementation in PMDK, Puddles,\ignore{ using undo-logging,}
\added{and Romulus.} Both the keys and the values are 8
bytes. Similar to the Linked List benchmark, \xname perform as fast as or better
than PMDK across the three operations \added{while being competitive with
  Romulus.} In summary, \xname{}' native-pointer results in a much faster
(\speedupbtreepointerchase\x{}) performance over PMDK for search operations.

\nfigure[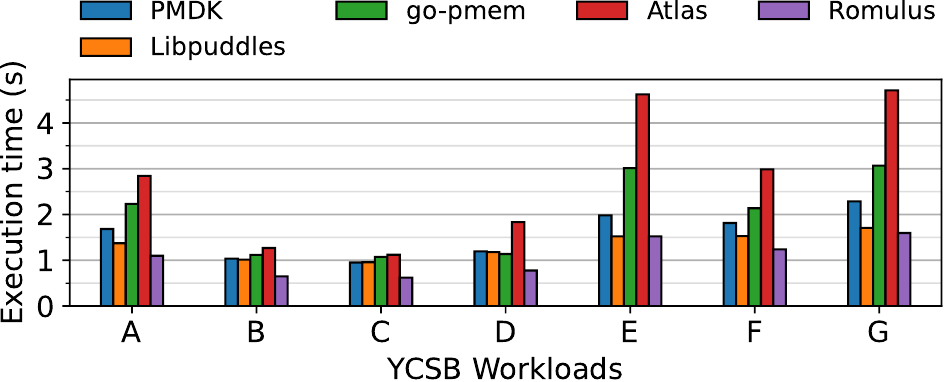,{KV Store implementation using different PM
    programming libraries, evaluated using YCSB
    workloads.},fig:kv-store-ycsb]

  \simplepara{KV-Store} To \added{evaluate} \xname{}' performance in databases,
  we evaluate PMDK's Key-Value store using \xname (using undo logging), PMDK,
  Atlas~\cite{atlas}, go-pmem~\cite{george2020go}, and
  Romulus~\cite{correia2018romulus}.  \reffig{fig:kv-store-ycsb} shows the
  performance across these libraries using the YCSB~\cite{ycsb} benchmark. For
  each workload, we run a 1 million keys load workload followed by a run
  workload with 1 million operations. Across the workloads, \xname are at least
  as fast and up to \maxspeedupycsb{}\x{} faster than PMDK. \added{Against
    Romulus, \xname is between 36\% slower to being equally fast across the YCSB
    workloads. Romulus's performance improvement is from its use of DRAM for
    storing crash-consistency logs. {While \xname' implementation is slower than
      Romulus, \xname's relocation and native pointer support is compatible with
      in-DRAM logs and could be used to improve its performance.}}

 \nfigure[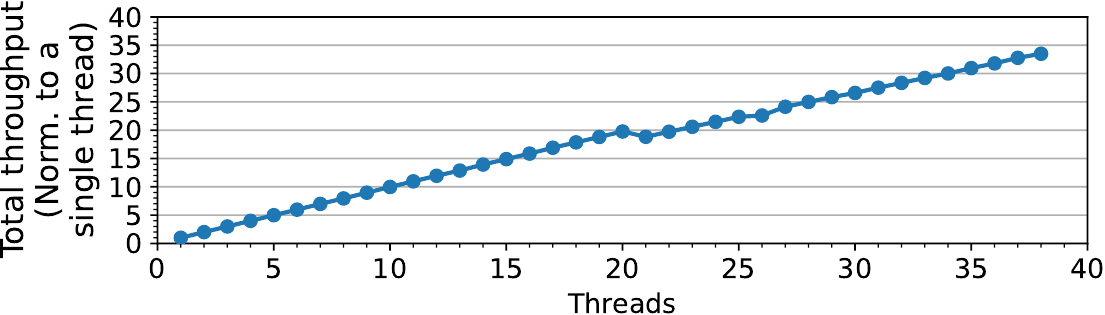,{Multithreaded workload that processes
   1/n$^\text{th}$ of the array per thread. \ignore{\xname{}' asynchronous logging
   interface allows it to have fast and scalable
   transactions.}},fig:euler-throughput]

\simplepara{Multithreaded scaling.}\label{par:multithreading} To study the
multithreaded scalability of \xname, we used an embarrassingly parallel workload
that computes Euler's identity for a floating-point array with a million
elements.  \reffig{fig:euler-throughput} shows the normalized time taken by the
workload with the increasing thread count scales linearly and is not limited by
\xname' implementation. \hl{In the benchmark, each worker thread works on a
  small part of the array at a time using a transaction.  The workload's
  throughput scales linearly with the number of threads until it uses all the
  physical CPUs (20); increasing the number of threads further still results in
  performance gains, albeit smaller. \xname{}' asynchronous logging interface,
  along with thread-local transactions, allows it to have fast and scalable
  transactions.}

\subsection{Relocation: Sensor Network Data Aggregation}\label{subsec:relocation-exp}

\nfigure[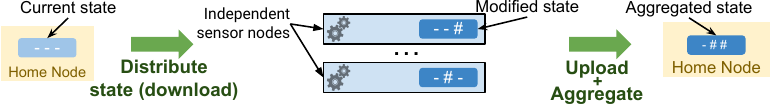,{\added{\textbf{Data Aggregation Workload.}
    Independent sensor nodes modify copies of pointer-rich data-structures and a
    home node aggregates the copies into a single copy.}},fig:data-agg]

\xname{}' ability to relocate data allows it to merge copies of PM data without
performing expensive reallocations or serialization/de-serializations. In
contrast, applications using traditional PM libraries cannot clone and open
multiple copies of PM data because they contain embedded UUIDs or virtual memory
pointers.

\added{To demonstrate the ability to relocate PM data across machines, we model
  a sensor network data-aggregation workload that combines several copies of PM
  data structures together. \reffig{fig:data-agg} shows the processing pipeline
  for this workload. A home node copies a PM-data structure to multiple
  independent sensor nodes that have their own puddle space. The independent
  nodes modify these copies and upload the result back to the home node which
  aggregates the states into a single data structure. Each node modifies the
  state data using \xname' transactions and can crash during writes. To model
  independent nodes with isolated persistent address spaces, we run the nodes in
  isolated docker containers.}

\added{\xname' ability to resolve address space conflicts in PM data and support
  for aggregating data allow the nodes to export their state as a portable
  format to the file system. The home node aggregates the states by reopening
  the data from each node, and \xname{} seamlessly rewrite all the pointers to
  make the data available for access. PMDK, on the other hand, does not support
  reading multiple copies of the same data within a single process. For the home
  node to aggregate the state, it needs to open each copy sequentially and
  reallocate the data into a larger pool.}

\begin{figure}
  \includegraphics[width=\linewidth]{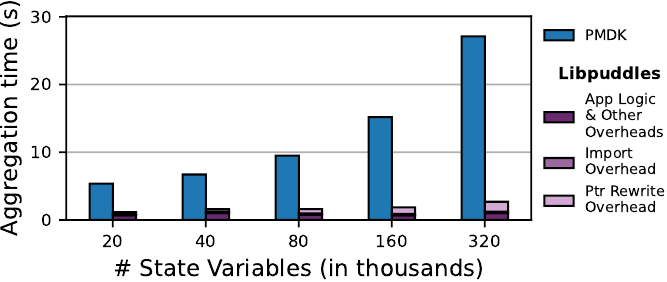}
  \caption{Total time taken by PMDK and \xname{} to aggregate PM data from 200
    sensor nodes\ignore{ with a varying number of state variables}.}
  \label{fig:form-demo}
\end{figure}

\added{\reffig{fig:form-demo} shows the total time spent and \xname' break down
  while aggregating states from 200 nodes with 100 to 1600 state variables
  each. Since PMDK needs to reallocate all the data, it is between
  \speedupformdemo\x{} to \maxspeedupformdemo\x{} slower than \xname{}. For
  \xname, the aggregation has a constant import overhead of 0.2~s, while pointer
  rewrite overhead scales with the number of elements and increasingly dominates
  the execution.}

\section{Related Work}
\label{sec:related}

Prior persistent memory works have used a variety PM pointer formats; PM
Libraries often use non-standard pointer formats that require translation to
use~\cite{pmdk, corundum, nvheaps, bittman2020twizzler} or do not allow the
programmer to reference data across PM regions, e.g., pools~\cite{pmdk,
  corundum, nvheaps}, limiting PM programming flexibility. \added{Some
  persistent memory programming libraries like Pronto~\cite{pronto} simplify
  PM programming by semantically recording updates like linked list inserts
  instead of individual memory writes. Unlike \xname, Pronto and
  Romulus~\cite{correia2018romulus} use DRAM for the working copy of the
  application data.}

\ignore{Many logging PM programming libraries use Write Ahead Logging (WAL) to
  maintain crash consistency. WAL can be implemented using undo
  logging~\cite{atlas, george2020go} to roll the transaction back on failure,
  redo logging~\cite{gu2019pisces, mnemosyne, giles2013software}, to roll the
  transaction forward on failure, or on a combination of undo and redo
  logging~\cite{nvheaps, corundum, pmdk, chatzistergiou2015rewind} and use
  multi-staged transactions. \xname{}'s system support provides a flexible
  logging interface and can support any combination of redo and undo logging
  even within the same transaction. This allows \xname to support recovery for a
  large number of existing libraries to benefit from a unified system-supported
  logging interface.}

\ignore{Researchers have also proposed logging techniques that rely on
  copy-on-write or shadow logging~\cite{liu2017dudetm, castro2018hardware,
    pangolin, mod}, and use custom logging solutions, e.g., semantically logging
  persistent memory operations~\cite{pronto}. Further, some PM libraries
  re-execute or resume PM programs during recovery~\cite{clobbernvm,
    izraelevitz2016failure, liu2018ido} or trace reachable objects on
  recovery~\cite{cai2020understanding}. LSNVMM~\cite{lsnvmm} uses a
  log-structured file-system-like approach to crash-consistency by directly
  appending PM data to a log. While these logging techniques cannot be directly
  used with \xname, these systems can still be modified to provide support for
  \xname{} native-pointers and relocatability.}

Researchers have previously proposed having a global unified virtual memory
space that all applications allocate from~\cite{redell1980pilot,
  bittman2019tale, kale1997design,
  bittman2020twizzler}. TwizzlerOS~\cite{bittman2020twizzler} is one such system
for persistent memory that proposes a global persistent object space similar to
\xname{}. \xname differ from TwizzlerOS in three major ways: (a) recovery in
TwizzlerOS, like PMDK, relies on the application, (b) unlike native virtual
pointers in \xname, TwizzlerOS uses redirection tables and index-based pointers
that can have up to 2 levels of indirections, and (c) finally, unlike \xname{},
TwizzlerOS does not support exporting data structure out of its global object
space. However, \xname{}' recovery and relocation support are orthogonal to
TwizzlerOS and can be implemented as an extension to TwizzlerOS. While
TwizzlerOS offers a new PM model, the open-source version does not support
crash-consistent allocations, making meaningful comparison impossible. And thus,
we do not evaluate TwizzlerOS against \xname{}.

Similar to TwizzlerOS, several previous OS works have looked into using a single
per-node unified address space. Opal~\cite{chase1992opal},
Pilot~\cite{redell1980pilot}, and SingularityOS~\cite{hunt2007singularity} all
provide a single address space for all the processes in a system. While OS like
Opal support single, unified address space with the ability to address
persistent data, they still suffer from the same limitations that today's PM
solutions do.

{Opal, for example, offers a global persistent address space, yet it lacks
  consistency or location independence.  \ignore{Opal, for example, provides a
    global persistent address space but does not provide Consistency or Location
    independence.} Data in Opal is inconsistent until a PM-aware application
  with write permissions reads it. \xname, on the other hand, guarantee
  system-supported recovery with no additional cost other than the one-time
  setup overhead. Further, since Opal has no information about the pointers
  embedded in the data, like PMDK, it requires expensive
  serialization/deserialization to replicate data structures within the address
  space. \hl{No support for pointer translation also means that Opal cannot
    relocate data structures on an address conflict when importing data from a
    different address space.}}

GVMS~\cite{heiser1993mungi} also introduces the idea of a singular global
address space, but for all the application data and shares it across multiple
cluster nodes to provide shared memory semantics. In contrast, \xname{} provide
a unified address space only for PM while still using traditional address spaces
for isolation and security.

Hosking \etal{}~\cite{hosking1993object} present an object store model
\added{for SmallTalk} that maps objects missing from the process address space
on a page fault similar to \xname{}\added{, but relies on SmallTalk's runtime
  indirection for checking and rewriting pointers. Moreover, their solution does
  not allow storing native pointers in storage, requiring translating pointers
  every time persistent data is loaded.} \added{In contrast, \xname does not
  depend on a specific runtime for identifying pointers, and provides
  application independent recovery and location independence.}\impr{Talk about
  the persistent object stuff in more detail and how they are different from
  libpuddles.}

  \added{Wilson and Kakkad et al.~\cite{wilson1991pointer,kakkad1999address}
    propose pointer translation at page fault time similar to \xname, however,
    their solution suffers from several problems. One of the major limitations is
    no support for objects that span multiple pages as each page can be
    relocated independently, breaking offset-based access into the
    object. \xname solve this problem by translating pointers at puddle
    granularity, allowing objects to span pages. Further, their solution does
    not support locating pointers in persistent data, and unlike \xname' pointer
    maps, they leave it as future work. Finally, unlike direct-access (DAX)
    support in \xname, their solution requires mapping data to the page cache as
    the data is stored in a non-native pointer format.}

\section{Conclusion}
\label{sec:conclude}

Current PM programming solutions' choices introduce several
limitations that make PM programming brittle and inflexible. They fail to
recover PM data to a consistent state if the original application writing the PM
data is no longer available or if the user no longer has write permission to the
data.  Existing PM systems also non-optimally choose among pointer choices that
result in unrelocatable PM data and, in some cases, performance overhead.

We solve these problems by providing a new PM programming library--\xname that
supports application-independent crash recovery and location-independent
persistent data.  To support this, \xname{} register logs with the trusted
daemon that manages and allocate persistent memory and automatically replays
logs after a crash.  The puddle system has a single global PM address space that
every application shares and allocates from.  A global address space and PM data
relocation support allows the use of native, unadorned pointers.

\ignore{These features allow \xname{} to simplify PM programming interface while
providing better performance than existing PM programming solutions.  \xname{}'
native virtual pointers provide a significant performance improvement over
PMDK's fat pointers.  Moreover, \xname supports the ability to relocate PM data
seamlessly and faster than traditional solution.}

\xname{}' native virtual pointers provide a significant performance improvement
over PMDK's fat pointers.  Moreover, \xname support the ability to relocate PM
data seamlessly and faster than traditional solutions.

\section*{Acknowledgement}
This work was supported in part by the ACE Center for Evolvable Computing, one
of the seven centers in JUMP 2.0, a Semiconductor Research Corporation (SRC)
program sponsored by DARPA.
\bibliographystyle{plain}
\bibliography{libpaper/common,libpaper/nvsl,paper}

\newpage
\appendix
\section{Artifact Appendix}

\newcommand{\code}[1]{\texttt{\footnotesize{}#1}}

\subsection{Abstract}
Puddles is a new persistent memory programming abstraction that provides support
for PM data that is easily mappable into a process address space, shareable
across processes, shippable between machines, consistent after a crash, and
accessible to legacy code with fast, efficient pointers as first-class
abstractions.

Provided artifact aims to verify the performance claims in the paper's Figure 9,
10, and 11.

\subsection{Description \& Requirements}
\subsubsection{How to access}
The artifact is available on Zenodo: \url{https://doi.org/10.5281/zenodo.8400339}

\subsubsection{Hardware dependencies}
The artifact requires and x86-64 machine and at least one Intel Optane DC-PMM
mounted at \code{/mnt/pmem0/}.

\subsubsection{Software dependencies}
\begin{enumerate}
  \begin{multicols}{2}
  \item Ubuntu 20.04\footnote{with no existing PMDK installation}
  \item g++ and gcc v10
  \item clang and clang++ v10
  \item libboost-system-dev
  \item libarchive-dev
  \item cmake
  \item make
  \item pmdk-tools
  \item git
  \item autoconf
  \item libpmemobj-dev
  \item bsdmainutils
  \item python3-pip
  \item libz-dev
  \item pkg-config
  \item expect
  \item wget
  \item golang-go
  \item llvm
  \end{multicols}
\end{enumerate}

\subsubsection{Benchmarks}
The artifact uses the following benchmarks: (a) YCSB (b) Examples from PMDK.

\subsection{Set-up}
\begin{enumerate}
\item The artifact requires enabling userfaultfd:\\
  \code{\footnotesize{}echo 1 | sudo tee /proc/sys/vm/unprivileged\_userfaultfd}
\item Optane DC-PMM should be mounted at \code{/mnt/pmem0} with
  \code{dax} support. This can be checked using:\\
  \code{\footnotesize{}mount | grep /mnt/pmem0}
\end{enumerate}

\subsection{Evaluation Workflow}
\subsubsection{Using Docker Containers}
The artifact can be run inside a docker container but needs access to host's \code{/mnt/pmem0} and \code{/dev/shm}.

\vspace{0.2cm}
\begin{mdframed}[backgroundcolor=Black!10, roundcorner=10pt,leftmargin=1, rightmargin=1, innerleftmargin=5, innertopmargin=1,innerbottommargin=5, outerlinewidth=2, linecolor=black]
  \begin{lstlisting}
# Build the docker image and the artifact
docker build -t libpuddles .

# Start the docker container
docker run --privileged  -v /mnt/pmem0/:/mnt/pmem0 --ipc=host -it libpuddles:latest

# Run evaluation from inside the docker container
./setup-and-run.sh --skip-deps
  \end{lstlisting}
\end{mdframed}
\vspace{0.2cm}

\subsubsection{Using a Bare Metal Machine}
To set up dependencies and run the benchmark, execute the following script:

\vspace{0.2cm}
\begin{mdframed}[backgroundcolor=Black!10, roundcorner=10pt,leftmargin=1, rightmargin=1, innerleftmargin=5, innertopmargin=1,innerbottommargin=5, outerlinewidth=2, linecolor=black]
  \begin{lstlisting}
./setup-and-run.sh
  \end{lstlisting}
\end{mdframed}
\vspace{0.2cm}
\begin{mdframed}[backgroundcolor=Black!10, roundcorner=10pt,leftmargin=1, rightmargin=1, innerleftmargin=5, innertopmargin=5,innerbottommargin=5, outerlinewidth=2, linecolor=black]
 \textbf{NOTE}: To skip dependency installation, the \\\code{setup-and-run.sh} script can be passed an optional \code{--skip-deps} argument:\\
 \code{./setup-and-run.sh --skip-deps  \# Run evaluation only}
\end{mdframed}
\vspace{0.2cm}

Although libpuddles is automatically compiled by this script, to manually compile it, checkout\\\code{artifact-root/libpuddles/README.md}.

\subsubsection{Major Claims}
The artifact provided reproduces the following claims from the paper:

\begin{enumerate}
\item \textbf{Figure 9}. Puddles’ performance comparison against PMDK and Romulus for singly linked list.
\item \textbf{Figure 10}. Performance comparison of Puddles, PMDK, and Romulus’s implementation of an order 8 Btree.
\item \textbf{Figure 11}. KV Store implementation and performance comparison using different PM programming libraries, evaluated using YCSB workloads.
\end{enumerate}

\subsubsection{Experiments}
The script \code{./setup-and-run.sh} runs three experiments to reproduce the
three major claims. To manually run them, run the following commands:

\vspace{0.2cm}
\begin{mdframed}[backgroundcolor=Black!10, roundcorner=10pt,leftmargin=1, rightmargin=1, innerleftmargin=5, innertopmargin=1,innerbottommargin=5, outerlinewidth=2, linecolor=black]
  \begin{lstlisting}
# Figure 9
libpuddles-scripts/run/linkedlist.sh
./plot-fig9.py

# Figure 10
libpuddles-scripts/run/btree.sh
./plot-fig10.py

# Figure 11
libpuddles-scripts/run/simplekv.sh
./plot-fig11.py
  \end{lstlisting}
\end{mdframed}
\vspace{0.2cm}

\subsection{Notes on \code{libpuddles-scripts/run/*.sh} Scripts}
Each of these scripts automatically starts an instance of the puddle daemon
(\code{puddled}), cleans up any persistent files, and runs a specific
workload.

These scripts, however, rely on certain environment variables:
\begin{enumerate}
\item \code{LIBPUDDLES\_PUDDLED\_PORT}: Port to use for the puddled daemon
\item \code{LIBPUDDLES\_DIR}: Directory with libpuddles' source (\code{artifact-root/libpuddles} for this artifact)
\item \code{ROMULUS\_DIR}: Directory with Romulus's source \\(\code{artifact-root/OneFile-romulus} for this artifact)
\end{enumerate}

\subsection{Execution Flow (\code{setup-and-run.sh})}
The \code{setup-and-run.sh} script automatically runs the following steps:

\begin{enumerate}
\item Setup the environment variables:
  \begin{enumerate}
    \item \code{LIBPUDDLES\_PUDDLED\_PORT}
    \item  \code{LIBPUDDLES\_DIR}
    \item  \code{ROMULUS\_DIR}
  \end{enumerate}
\item Install all dependencies.
\item Compile all sources.
\item Enable userfaultfd:\\
  \code{\footnotesize{}echo 1 | sudo tee /proc/sys/vm/unprivileged\_userfaultfd}
\item \textbf{Run experiments}
\item Run \code{libpuddles-scripts/run/linkedlist.sh} and write results to \code{libpuddles-scripts/data/linkedlist.csv}.
\item \textbf{Plot Figure 9} using the linkedlist data and write to \code{plots/ae/fig9.pdf} and \code{plots/ae/fig9.pdf}
\item Run \code{libpuddles-scripts/run/btree.sh} and write results to \code{libpuddles-scripts/data/btree.csv}.
\item \textbf{Plot Figure 10} using the btree data and write to \code{plots/ae/fig10.pdf} and \code{plots/ae/fig10.pdf}
\item Run \code{libpuddles-scripts/run/simplekv.sh} and write results to \code{libpuddles-scripts/data/simplekv.csv}.
\item \textbf{Plot Figure 11} using the simplekv data and write to \code{plots/ae/fig10.pdf} and \code{plots/ae/fig11.pdf}
\end{enumerate}

\end{spacing}

\ifannotated
  {
    \newpage
    \setcounter{section}{0}
    \title{Additional Material}
    \author{}
    \date{}
    \renewcommand{\addresses}{}
    \maketitlesup
    \input{additional-material}
  }
\fi

\end{document}